\documentclass[aps,pra,preprint,amsmath,mathtools,amssymb,longbibliography]{revtex4}
\usepackage{graphicx,epsfig,epsf,color,hhline}
\usepackage{bm}
\usepackage{natbib}
\usepackage{braket,ulem}
\usepackage{amsmath}

\usepackage[T1]{fontenc}
\usepackage[utf8]{inputenc}

\usepackage{physics}

\begin{document}
	
\def\vec#1{\bm{#1}}

\title{Quantum-router: Storing and redirecting light at the  photon level} 
\author{Martin C. Korzeczek and Daniel Braun}
\affiliation{Eberhard-Karls-Universit\"at T\"ubingen, Institut f\"ur Theoretische Physik, 72076 T\"ubingen, Germany}

\centerline{\today}

\begin{abstract}
  We propose a method for spatially re-routing single photons or 
  light in a coherent state with small average photon number 
  by purely electronic means, 
  i.e.~without using mechanical devices such as micro-mirror arrays. 
  The method is based on mapping the quantum state
  of the incoming light onto 
  a spin-wave in an atomic ensemble
  as is done in quantum memories of light.
  Then the wavevector of the spin-wave is modified in a controlled way 
  by an applied magnetic field gradient. 
  Finally, by re-applying the same control beam as for storing, 
  the signal pulse is released in a new direction 
  that depends on the deflected wavevector of the spin-wave. 
  We show by numerical simulation that efficiencies can be achieved 
  for arbitrary deflection angles in the plane 
  that are comparable with simple photon storage and 
  re-emission in forward direction, 
  and propose a new method for eliminating the stored momentum 
  as source of decoherence in the quantum memory. 
  In a reasonable parameter regime, 
  the re-routing should be achievable on a time-scale 
  on the order of few to $\sim100$ microseconds, depending on the deflection angle.
  The shifts in the wavevector that can be achieved using the Zeeman-effect, 
  with otherwise minimal changes to the spin-wave,
  can also be used to complement existing ac-Stark spin-wave manipulation methods.
\end{abstract}
\maketitle

\setcounter{page}{0}
\pagenumbering{arabic}

\section{Introduction}
Light is a natural carrier for information, both classical and
quantum, due to its large speed, relatively weak interaction with
matter, and the possibility to guide light through optical fibers.
The weak interaction motivates, on the other hand, to develop
light-matter interfaces, such that quantum information can be
stored and processed in other systems.  It is well known that the
efficiency with which light can be stored in matter can be increased
by using an ensemble of atoms.  The coupling constant relevant for the
absorption of a single photon increases then $\propto\sqrt{N}$ with the
number $N$ of atoms. It is nevertheless challenging to coherently
absorb, store, and release again a single photon with an ensemble of
atoms.  A number of techniques have been developed to that end over
the years such as  
EIT, slow light (for a review see \cite{lvovsky_optical_2009}),
controlled reversible inhomogeneous broadening (CRIB)
\cite[and 14-15 therein]{sangouard_analysis_2007}, 
and atomic frequency combs (AFC).  
In the latter, the  distribution of atomic
density over detuning has a comb-like structure, leading to multimode capacity.
Even photon pairs have been coherently stored and released again,
keeping part of their initial entanglement \cite{tiranov_temporal_2016}, 
as required by the DLCZ protocol of entanglement swapping for long distance
quantum communication \cite{duan_long-distance_2001}.  
A basic working principle of these memory schemes 
is the storage of phase information of the incoming mode in a
collective atomic excitation, such as a spin-wave, where each atom
contributes part of the excitation with a well defined phase.  
Ideally, the phase relations remain intact during the storage time, a
requirement that can be achieved to a high degree by using hyperfine
spin states that decohere very slowly.\\

Most of the previous work has focused on improving the storage of the
photon as measured by fidelity, bandwidth, 
and storage time or realizing quantum operations and mode multiplexing.
In the present work we are interested in another aspect: the control
of directionality of the emitted pulse. As was noted in
\cite{scully_directed_2006}, the phases of the individual atomic
contributions in the spin wave are such that the signal is re-emitted
in exactly the same direction as in which it was absorbed. This gives
the intuition that the directionality for collective emission
is encoded in Hilbert-space phases and can be controlled
by manipulating these phases prior to emission. 
Indeed, from
\cite{scully_directed_2006} it is 
clear that if one created phases that correspond to those that would have
resulted from absorption from a different direction, re-emission would be
in that direction.

The importance of the phases during re-emission has been considered before:
Chen et al.~\cite{chen_coherent_2013} 
demonstrated forward and backward retrieval with EIT. 
Backward retrieval can lead to higher fidelity due to 
reduced re-absorption and compensation of the Doppler shift. 
In \cite{sangouard_analysis_2007} it was
noted that by suitably changing the 
phases, the signal is re-emitted in backward direction compared to the original
incoming signal without the need of additional control lasers.
In \cite{chen_controllably_2016, wang_three-channel_2009}, forward retrieval and routing with a
small 'array' of possible control beams was achieved. 
\cite{surmacz_efficient_2008} recognized phase matching and
the spin-wavevector $\kappa$
as important for directionality and proposed multi-mode storage by
having an array of control fields with sufficiently 
differing angles that any control beam only affects its own spin wave.
As noted in \cite{mazelanik_coherent_2019}, 
imprinting a position-dependent phase $e^{i\phi(\vec{r})}$ 
onto the atomic coherence has, in $k$-space, 
the effect of a convolution of the original spin-wave and the added phase-factors. 
Due to the condition of phase matching, the $k$-space 
contributions of the spin-wave define whether 
and in which direction the signal will be re-emitted 
upon arrival of the next control pulse. 
Added phases that are linear in position shift the wavevector 
stored in the spin wave \cite{leszczynski_spatially_2018}, 
periodic phases will coherently divide the spin-wave 
into several contributions with shifted wavevectors 
\cite{parniak_quantum_2019, mazelanik_coherent_2019}.   
Ref.~ \cite{leszczynski_spatially_2018, parniak_quantum_2019, 
	mazelanik_coherent_2019} 
proposed and demonstrated experimentally
the use of an ac-Stark shift for manipulating the spin wave as described above, 
implementing temporal as well as directional beam splitters, 
and observing the Hong-Ou-Mandel effect. 
In \cite{lipka_spatial_2019}, the ac-Stark effect is demonstrated 
to allow for mimicking the effect of a cylindric lens 
by imprinting phases $\propto y^2$ orthogonal to the emission direction. 
The ac-Stark shift is thus a powerful tool for 
coherently manipulating spin waves. 
Solely shifting the wavevector of the spin waves by a large amount, 
thus changing the emission direction without splitting the spin wave 
is hard to achieve using this method, 
as inducing a suitable energy shift linear in space 
over the whole atomic cloud requires
correspondingly large absolute shifts at some part of the cloud.
\cite{mazelanik_coherent_2019} report an ac-Stark induced energy shift 
on the order of MHz for 0.1\,W laser power, while a magnetic field 
creates $\sim$ 10\,MHz per Gauss, 
such that the shift can reach the GHz regime.

\begin{figure}
	\centering
	\includegraphics[width=1.\linewidth]{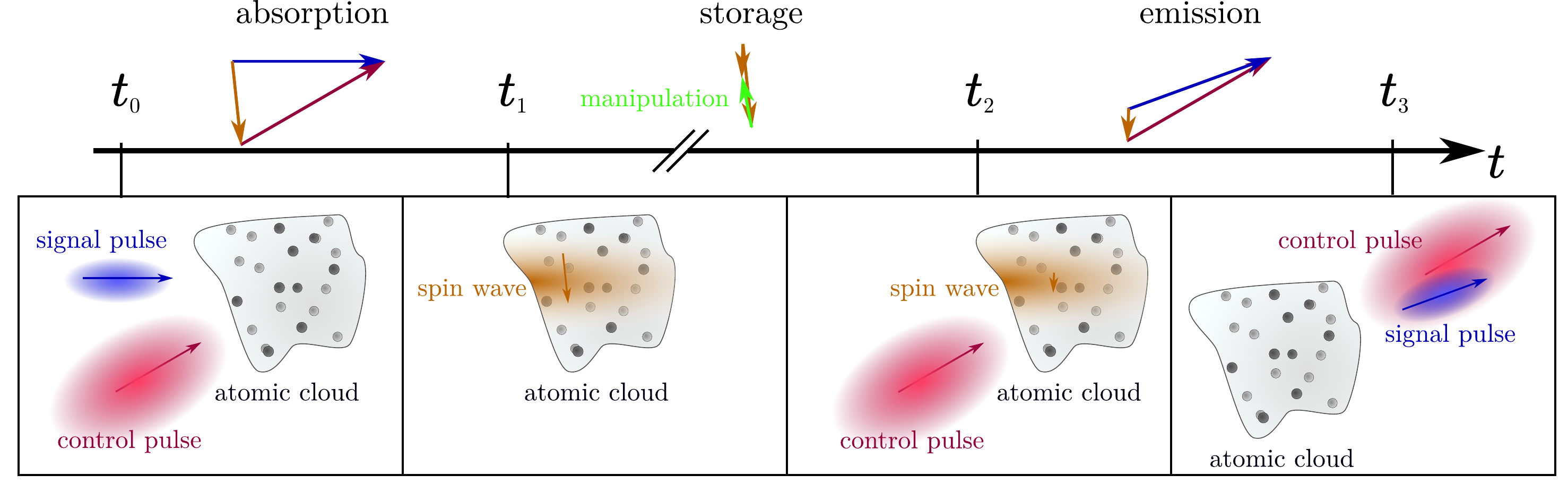}
	\caption{The time line is divided into the stages of
          absorption, storage, and emission. For each stage, the
          relevant wavevectors for phase matching are drawn above the
          axis, and a depiction of the system's state at the beginning
          and end of each stage is shown below. A ``manipulation''
          (momentum change of the spin-wave) during the storage phase
          allows re-emission in a new direction.}
	\label{fig:timeline}
      \end{figure}

Here we extend these previous works to allow
emission in an arbitrary direction in the 2D plane by manipulating the spin-wave
phases in a controlled way during the storage phase
(see Fig.~\ref{fig:timeline} for a schematic description 
of the pulse sequence).  
We show that in doped solids, 
where the atoms carrying the spin-wave can be 
considered to sit at fixed positions,
this can be achieved 
by applying a magnetic field gradient and using the Zeeman effect 
for reasonable coil parameters and power supplies.
This allows for fast routing of photons (few to $\sim100\,\mu$s 
with reasonable parameters, depending on the deflection angle) 
without using any mechanical
parts, i.e.~the re-emission direction is controlled by purely electronic means. 
Even without optimizing the parameters of the control beam, efficiencies
of the re-emission in any direction can be achieved that are 
comparable to those of forward re-emission. 
In cold atomic clouds or hot atomic vapors, where atomic motion scrambles the phases of spin waves that carry significant momentum, deflection angles up to $\sim 20\,$mrad
and $\sim 0.2\,$mrad, respectively, should be achievable, 
which still allow for fast photon routing.

Given the role of the individual atomic phases 
and the ability to shift $\kappa$ in the spin-wave, 
we also propose a new way of avoiding
decoherence due to the interaction of diffusion and the momentum stored in the spin-wave. This can contribute to relax the necessity of using co-propagating pulses in implementations where the atoms move freely. 

The dominant decoherence mechanism in Raman-type 
quantum-memories is ground state decoherence.  In vapor cells, it
results mostly from the drift of atoms in and out of the laser beam, and
in ultracold gases often from uncontrolled magnetic fields \cite{lvovsky_optical_2009}. 
In the latter case, an improvement can be obtained by using atomic clock states  \cite{chen_controllably_2016,zhao_millisecond_2009} 
(i.e.~states with a transition frequency which is constant 
to first order in changes to the magnetic field), 
in the former by
using optical lattices for limiting the motion of the atoms. Using rubidium, storage times reaching $1/e$ lifetimes of $0.22\,$s \cite{yangEfficientQuantumLight2016} for single light quanta and $16\,$s \cite{dudinLightStorageTime2013} for coherent states were reported. Using dopants in a solid, \cite{maOnehourCoherentOptical2021a} report storage times over $1\,$h. 
Reviews over different approaches to quantum memories are \cite{heshami_quantum_2016,lvovsky_optical_2009}.

\section{The System} 
\label{sec:System}
The system consists of an atomic cloud with atomic density $n(\vec{r})$ 
and a total of $N$ atoms inside of a geometrical volume $\mathcal{V}$ 
with $\mathop{\mathrm{Vol}}(\mathcal{V})=V$.  
Three internal states $\ket{g},\ket{e},\ket{s}$ in $\Lambda$-configuration 
are taken into account, 
and the motional state $\ket{\psi}$ is given by a wave function 
$\psi(\vec{r}_1,\dots,\vec{r}_N)$ which is a product 
of single-particle wave packets. 
We assume the atoms to be localized on a scale much smaller 
than the photonic wave lengths. In experiment, this can be realised using warm vapors, cold atomic clouds, or dopants inside a solid body.
With this, averaging over radius-$\epsilon$ spheres $v_{\vec{r}}$ 
around position $\vec{r}$ much smaller than the wave lengths 
and much bigger than the atomic wave functions 
allows for introducing the atomic density $n(\vec{r})$ 
as the approximate eigenfunction of the atomic density operator 
averaged over the spheres $v_{\vec{r}}$:
\begin{eqnarray} \label{eq:n}
\hat{n}({\vec{r}}) \ket{\psi} := \left(\sum_{i=1}^N 
      		\frac{1}{\mathop{\mathrm{Vol}}(v_{\vec{r}})}\ket{v_{\vec{r}}}_i\!\bra{v_{\vec{r}}}\right) 
      		\ket{\psi} \approx n({\vec{r}}) \ket{\psi},
\end{eqnarray} 
where $\ket{v_{\vec{r}}}_i\!\bra{v_{\vec{r}}}:=
      	\int_{v_{\vec{r}}} d^3 r'\ \ket{\vec{r}'}_i\!\bra{\vec{r}'}$.

\begin{figure}
	\centering
	\includegraphics[width=.6\linewidth]{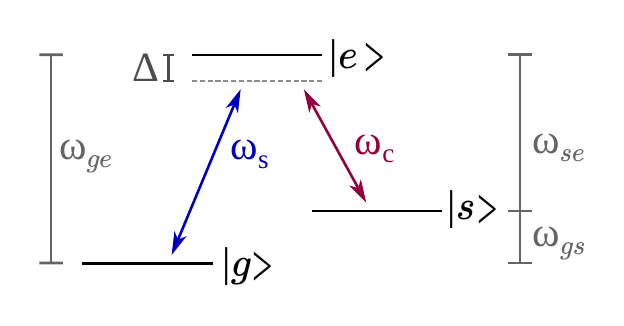}
	\caption{The energy levels of the atoms and relevant notation.\label{fig:energy_levels}}
\end{figure}

The atoms are treated as frozen in place for the absorption and emission processes.
The definitions and derivations are parallel to the ones introduced 
in \cite{gorshkov_universal_2007,gorshkov_photon_2007-2}, 
and modified for 3d-space with arbitrary signal and control directions, 
as well as the quantized atomic motional state given above. 
Detailed derivations and outline of the numerical procedure are given in \cite{master_thesis}. 
Atomic transition operators for atom $i$ 
are denoted by $\hat{\sigma}_{\mu\nu}^i=\ket{\mu}_i\!\bra{\nu}$ ($\mu,\nu\in \{e,s,g\}$) 
and couple to the corresponding light modes via dipole transitions 
as depicted in Fig.~\ref{fig:energy_levels}. 
The control field (index ``c'') is described classically 
by its positive frequency envelope $\mathcal{E}_\text{c}^{\vec{k}_\text{c}}(\vec{r},t)$. 
As in \cite{gorshkov_photon_2007-1}, the control pulse's influence 
on the atomic cloud is later described by 
half the induced Rabi frequency $\Omega(\vec{r},t)$ which will be defined shortly:
\begin{eqnarray}\label{eq:E_c}
	\vec{E}_\text{c}(\vec{r},t) &= \frac{1}{2} \vec{\epsilon}_{\text{c}} 
				e^{i (\vec{k}_\text{c}\cdot\vec{r}-c|\vec{k}_\text{c}|t)} 
				\mathcal{E}_\text{c}^{\vec{k}_\text{c}} + c.c.,
\end{eqnarray}
here $\vec{E}_\text{c}$ is the electric field of the control pulse, 
$\vec{\epsilon}_\text{c}$ its polarisation, 
$\vec{k}_\text{c}$ its dominant wavevector, $c$ is the vacuum speed of light, 
and $c.c.$ stands for the complex conjugate.
The signal pulse (index ``s'') is taken as fully quantised in 3d space 
with electric field operator
\begin{eqnarray}
\hat{\vec{E}}_\text{s}(\vec{r}) =& \sqrt{\frac{\hbar c}{2 \epsilon_0 (2\pi)^3}} 
			\sum_{\ell\in\{1,2\}} \int_{\vec{k}\in \mathbb{R}^3} d^3 k\ \sqrt{|\vec{k}|} 
			\vec{\epsilon}_{\vec{k},\ell} e^{i \vec{k}\cdot \vec{r}} \hat{a}_{\ell}(\vec{k}) + h.c., 
\end{eqnarray}
where $\epsilon_0$ is the electric vacuum permittivity, 
$h=2\pi\hbar$ is Planck's constant, 
$\vec{\epsilon}_{\vec{k},\ell}$ is the polarisation vector for polarisation $\ell$ 
and wavevector $\vec{k}$ and $\hat{a}_\ell (\vec{k})$ is 
the continuous-mode annihilation operator for polarisation $\ell$ 
and wavevector $\vec{k}$ with $\left[\hat{a}_{\ell}(\vec{k}),\hat{a}^{\dagger}_{\ell'}(\vec{k}')\right]= 
\delta(\vec{k}-\vec{k'}) \cdot \delta_{\ell,\ell'}$ 
and $h.c.$ stands for the hermitian conjugate.

As with the control field, 
we define positive frequency envelopes also for 
the signal field ($\hat{\mathcal{E}}^{\vec{k}_\text{s}}(\vec{r},t)$), 
the $g\leftrightarrow e$-coherence ($\hat{P}^{\vec{k}_\text{c}(\vec{r},t)}$, the ``polarisation'') 
and the $g\leftrightarrow s$-coherence ($\hat{S}^{\vec{\kappa}}(\vec{r},t)$, the ``spin wave''), 
\begin{eqnarray}\label{eq:def_envelopes}
\hat{\mathcal{E}}^{\vec{k}_\text{s}}(\vec{r},t) &=& \sqrt{\frac{V}{(2\pi)^3}} 
		e^{-i(\vec{k}_\text{s}\cdot\vec{r}-c|\vec{k}_\text{s}| t)}
		\int_{\vec{k}\in \mathbb{R}^3} d^3 k\ \sqrt{\frac{|\vec{k}|}{|\vec{k}_\text{s}|}}\ 
		\frac{\vec{d}\cdot \vec{\epsilon}_{\vec{k}}}{\vec{d}\cdot \vec{\epsilon}_{\vec{k}_\text{s}}} 
		e^{i \vec{k}\cdot \vec{r}} \hat{a}({\vec{k}}),\\	\nonumber
\hat{P}^{\vec{k}_\text{s}}(\vec{r},t) &=&  \frac{\sqrt{N}}{n(\vec{r})} \sum\limits_{i=1}^{N} 
		e^{-i(\vec{k}_\text{s}\cdot \vec{r}-c|\vec{k}_\text{s}|t)} \hat\sigma_{ge}^{i} 
		\,\frac{\ket{v_{\vec{r}}}_i\!\bra{v_{\vec{r}}}}{\mathop{\mathrm{Vol}}(v_{\vec{r}})}, \\ \nonumber
\hat{S}^{\vec{\kappa}}(\vec{r},t) &=&  \frac{\sqrt{N}}{n(\vec{r})} \sum\limits_{i=1}^{N} 
		e^{-i((\vec{k}_\text{s}-\vec{k}_\text{c})\cdot \vec{r}-c(|\vec{k}_\text{s}|-|\vec{k}_\text{c}|)t)} 
		\hat\sigma_{gs}^{i} \,\frac{\ket{v_{\vec{r}}}_i\!\bra{v_{\vec{r}}}}{\mathop{\mathrm{Vol}}(v_{\vec{r}})}, 
		\quad \vec{\kappa}:=\vec{k}_\text{s}-\vec{k}_\text{c}\\ \nonumber
\Omega(\vec{r},t) &=& \Omega^{\vec{k}_\text{c}}(\vec{r},t)\qquad  
		=  \qquad \frac{1}{2\hbar} \vec{d}_\text{c}\cdot \vec{\epsilon}_\text{c} 
			\mathcal{E}^{\vec{k}_\text{c}}_\text{c}(\vec{r},t)
\end{eqnarray}
and the corresponding interaction Hamiltonian 

\begin{eqnarray} \label{eq:interaction_hamiltonian}
\hat{H}_\text{I}=&-\sum_{j=1}^{N}\hat{\vec{d}}_j\cdot 
		(\hat{\vec{E}}_\text{s} (\hat{\vec{r}}_j)+\vec{E}_\text{c}(\hat{\vec{r}}_j,t))\\
\approx&  -\sum_{j=1}^{N} \left[ \sqrt{\frac{\hbar c}{2\epsilon_0 (2\pi)^3}} 
		\int_{\vec{k}\in \mathbb{R}^3} d^3 k\ \sqrt{|\vec{k}|} 
		\left( \vec{d}\cdot\vec{\epsilon}_{\vec{k}} 
				e^{i \vec{k}\cdot \hat{\vec{r}}_j} \hat\sigma_{eg}^j \hat{a}(\vec{k})+h.c.\right) 
		+ \right.\nonumber \\
& \left. \qquad \qquad +\ \frac{1}{2} \vec{d}_\text{c}\cdot\vec{\epsilon}_\text{c} 
		e^{i (\vec{k}_\text{c}\hat{\vec{r}}_j-c|\vec{k}_\text{c}|t)} 
		\hat{\sigma}_{es}^j \mathcal{E}_\text{c}(\hat{\vec{r}}_j,t) +h.c.\right] \nonumber\\ \nonumber
\approx & -\hbar \int_{\mathcal{V}} d^3 r\ \left[ \sqrt{N}g \hat{P}^{\vec{k}_\text{s}}(\vec{r},t) 
		\hat{\mathcal{E}}^{\vec{k}_\text{s}}(\vec{r},t) 
		+ \hat{S}^{\vec{\kappa}}(\vec{r},t) \Omega(\vec{r},t) + h.c.\right] \hat{n}(\vec{r}).
\end{eqnarray}
Here, $g=\sqrt{c|\vec{k}_\text{s}|/(2\hbar \epsilon_0 V)} \vec{d}\cdot \vec{\epsilon}_{\vec{k}_\text{s}}$ 
is the single particle atom-light coupling, $\vec{d}$ is the dipole moment 
of the $g\leftrightarrow e$-transition, $\vec{d}_\text{c}$ the dipole moment of the
$s\leftrightarrow g$-transition and $\vec{\kappa}$ is the wave
vector difference between the pulses. The signal and control
field polarisations are chosen to be $\vec{e}_z$ and the
$\ell$-index is discarded. We correspondingly consider all
involved wavevectors in the $xy$-plane. This allows for arbitrary deflection angles in the plane without complications from a change in polarisation. As the use of opposite circular polarisations for the probe and signal pulses in the $\Lambda$ scheme can strongly simplify distinguishing the two pulses for co-propagating configurations, in many implementations it will be advantageous to use a scheme with circular polarisation instead. However, this restricts the spin wave manipulation to deflection angles that do not strongly depart from forward or backward emission, so that the state overlap to the original polarisation remains high.

The Rotating Wave Approximation is used and it is assumed 
that the signal pulse only couples to the $g\leftrightarrow e$-transition 
and similarly the control pulse with the $s\leftrightarrow e$-transition. 

Initially all atoms are in the ground state $\ket{g}$ and, 
as atomic motion is frozen, also the Doppler effect is neglected. 
Inhomogeneous broadening in the context of photon storage 
in an ensemble of atoms was considered in \cite{gorshkov_photon_2007-3}. 
The signal pulse is taken to be a weak coherent state with $|\alpha|^2\ll N$, 
with $|\alpha|$ the expectation value of the photon number. 

With these initial conditions, the fields 
$\mathcal{E}(\vec{r},t),\ P(\vec{r},t)$ and $S(\vec{r},t)$ can be defined 
as the system state's eigenvalues to the corresponding operators:
$\mathcal{E}\leftrightarrow\hat{\mathcal{E}}^{\vec{k}_\text{s}}(\vec{r},t)$, 
$P\leftrightarrow \hat{P}^{\vec{k}_\text{s}}(\vec{r},t)$ 
and $S\leftrightarrow \hat{S}^{\vec{\kappa}}(\vec{r},t)$. 
Given our initial conditions and the limit of weak signal pulses, 
the system's state remains an eigenstate to these operators for all times, 
thus enabling our description through the complex-numbered eigenvalues. 
Choosing $\alpha=1$, all results for $\mathcal{E}$, $P$ and $S$ 
for a coherent signal pulse coincide with the expectation values of the operators 
that would result from using a 1-Photon Fock state as signal pulse. 
Therefore, 1-Photon Fock states can be described with the exact same formalism. 

The time evolution of the fields is given by 
the Heisenberg equation of motion and results in 
\begin{eqnarray}\label{eq:pde}
\left( \partial_t+c\partial_{\vec{e}_{\vec{k}_\text{s}}} \right) \mathcal{E}\approx& i \sqrt{N} g \frac{V}{N} n P,\\ \nonumber
\partial_t P =& -(\gamma+i \Delta) P + i \Omega S+i \sqrt{N} g \mathcal{E}, \\ \nonumber
\partial_t S =& i \Omega^* P,
\end{eqnarray}
where $\partial_{\vec{e}_{\vec{k}_\text{s}}}$ is a spatial derivative 
in direction $\vec{e}_{\vec{k}_\text{s}}:=\vec{k}_\text{s}/|\vec{k}_\text{s}|$, 
the direction of propagation of the signal pulse. 
$\gamma$ is the spontaneous emission rate of the excited state 
(which is added heuristically to describe 
the most basic effect of spontaneous emission), and $\Delta$ the detuning. 

The number of photons in the signal field is given by
\begin{eqnarray}
\langle \hat{N}_\text{ph}\rangle \approx \frac{1}{V} \int d^3 r\ \mathcal{E}^*(\vec{r},t) \mathcal{E}(\vec{r},t),
\end{eqnarray}
and the number of excitations stored in the atomic cloud is
\begin{eqnarray}
\langle \hat{N}_{\ket{s}}\rangle \approx& \frac{1}{N} 
		\int_{\mathcal{V}} d^3 r\ n(\vec{r}) S^*(\vec{r},t)S(\vec{r},t), \text{ and} \\
\langle \hat{N}_{\ket{e}}\rangle \approx& \frac{1}{N} 
		\int_{\mathcal{V}} d^3 r\ n(\vec{r}) P^*(\vec{r},t)P(\vec{r},t),
\end{eqnarray}
respectively. With these, the time evolution of our state 
(neglecting atomic motion and decoherence) 
is fully described by the complex-valued fields $\mathcal{E}, P$ and $S$ 
and their time evolution (\ref{eq:pde}), 
with a direct mapping to the corresponding quantum state (for the atomic degrees of freedom):
\begin{eqnarray}\label{eq:field_op_eigenstate}
      \ket{\Psi^P_{S}(t)}&=& \int d^{3N}r \bigotimes_{i=1}^{N} c_i
         		\ \Big(\ket{g}_i   
               		+e^{i(\vec{k}_\text{s}\cdot\vec{r}_i-c|\vec{k}_\text{s}|t)}
                       	\frac{P({\vec{r}_i,t})}{\sqrt{N}}\ket{e}_i \\
         		\nonumber &&\qquad\qquad\qquad 
         				+ e^{i((\vec{k}_\text{s}-\vec{k}_\text{c})\cdot\vec{r}_i-c(|\vec{k}_\text{s}|-|\vec{k}_\text{c}|)t)}
                        \frac{S({\vec{r}_i,t})}{\sqrt{N}} \ket{s}_i \Big)\times\\ \nonumber
            && \times  \psi(\vec{r}_1,\dots,\vec{r}_N)\ket{\vec{r}_1,\dots,\vec{r}_N}.
\end{eqnarray}
Here, $c_i\approx 1$ are normalisation factors.

\section{Dynamics and directionality}
\label{sec:Dynamics}
We partition the system dynamics into three stages as depicted
in Fig.~\ref{fig:timeline}: From $t_0$ to $t_1$, the
absorption takes place. There, the atoms 
start in the ground state and the incoming 
signal and control pulses meet in the atomic cloud where a fraction $\eta_{abs}$ of the excitations
of the probe pulse is converted into the spin wave.
Between $t_1$ and $t_2$, the light remains 
stored and
we optionally manipulate the spin wave using the Zeeman
effect. During this time, a slow decay of the spin wave occurs
but which we neglect in most of this work. 
 	During storage, the control field is absent,
$\Omega(\vec{r},t)=0$. 
From time $t_2$ on, the emission control pulse arrives 
and releases the excitations stored in the spin wave 
into a new signal pulse with a possibly altered direction 
and remaining fraction of original excitations 
$\eta=\eta_\text{abs}\ \eta_\text{em}$.   	

We consider in the following a spherical sample with volume $V=L^3$ 
and constant density, and  change to unit-free coordinates 
by using $L$ as length scale, $1/\gamma$ as time scale, and defining the atomic
number density relative to the mean density, $\tilde{n}$: 
\begin{eqnarray}
\tilde{\vec{r}}:=\frac{\vec{r}}{L}, \ \tilde{t}:=\frac{t}{1/\gamma}, 
\ \tilde{n}:=\frac{n}{N/V}, \ \tilde{c}:=\frac{c}{\gamma L}.
\end{eqnarray}
The simplifying assumption of a uniform atomic density 
allows for numerically simple PDEs. 
A treatment of exact atomic positions 
can be found in \cite{asenjo-garcia_exponential_2017,manzoni_optimization_2018}. 

We define 
\begin{eqnarray}
\tilde{\Delta}:=\frac{\Delta}{\gamma}, \ 
\tilde{\Omega}:=\frac{\Omega}{\gamma}, \ 
\tilde{g}:= \frac{\sqrt{N} g}{\gamma}, \ 
\tilde{P}:=\tilde{n}P, \ 
\tilde{S}:= \tilde{n} S,
\end{eqnarray}
with $\tilde{c}$ the dimensionless speed of light, 
$\tilde{\Delta}$ the dimensionless two-mode detuning, 
$\tilde{\Omega}$ half the dimensionless Rabi frequency 
induced by the control-pulse, 
and $\tilde{g}$ the dimensionless enhanced coupling 
between the atoms and the signal pulse. 
The normalised polarisation $\tilde{P}$ 
and the normalised spin wave $\tilde{S}$ 
are zero outside of the atomic cloud, 
which allows for a more direct interpretation 
of their numerical values when plotted.  
We define the $x$-axis such that $\vec{k}_\text{s}=k_\text{s} \vec{e}_{x}$.
The partial differential equations (PDEs) are then
\begin{eqnarray}\label{eq:final_pde}
\left( \partial_{\tilde{t}}+\tilde{c} \partial_{\tilde{x}} \right) 
		\mathcal{E}({\tilde{\vec{r}},\tilde{t}})&=& 
				i \tilde{g} \tilde{P}({\tilde{\vec{r}},\tilde{t}}),\\
\partial_{\tilde{t}} \tilde{P}({\tilde{\vec{r}},\tilde{t}})&=&
        -(1+i\tilde{\Delta})\tilde{P}({\tilde{\vec{r}},\tilde{t}})
        +i\tilde{\Omega}({\tilde{\vec{r}},\tilde{t}})
	\tilde{S}({\tilde{\vec{r}},\tilde{t}})
        +i \tilde{g} \tilde{n}({\tilde{ \vec{r}}}) 
	\mathcal{E}({\tilde{\vec{r}},\tilde{t}}), \nonumber \\
\partial_{\tilde{t}} \tilde{S}({\tilde{\vec{r}},\tilde{t}})&=& 
 i \tilde{\Omega}^*({\tilde{\vec{r}},\tilde{t}}) 
	\tilde{P}({\tilde{\vec{r}},\tilde{t}}). \nonumber
\end{eqnarray}
The optical depth $d$ as defined in
\cite{gorshkov_photon_2007-2} is here given by $d=
\tilde{g}^2/\tilde{c}$ when using $L$ as 
length scale. If the cloud diameter is used as length scale
instead, and the cloud has spherical shape and constant
density we get $d'\approx 1.24\ d$.
We consider the ideal situation of no dephasing during
the storage time. 
In Sec.~\ref{subsec:mode_mismatch} we shortly 
discuss the dephasing-relevant aspects connected to 
the wavevector stored in the spin wave $\kappa$. 

\subsection{Phase matching conditions and directionality}
For the absorption process, each excitation in the signal pulse 
carries the wavevector $\vec{k}_\text{s}$ and, if absorbed, 
leads to the emission of a control field excitation 
with wavevector $\vec{k}_\text{c}$ such that a spin wave excitation 
with wavevector 
\begin{equation}
\vec{\kappa}=\vec{k}_\text{s}-\vec{k}_\text{c} \label{eq:phase_match_abs}
\end{equation}
remains due to the conservation of momentum.

\begin{figure}
	\centering
	\includegraphics[width=0.3\linewidth]{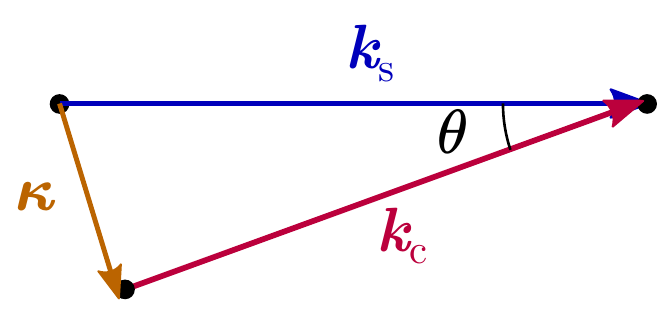}
	\hspace*{.1\linewidth}
    \includegraphics[width=0.3\linewidth]{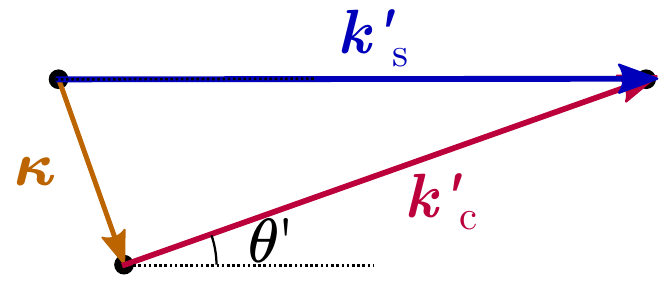}
	\caption{The phase matching condition for the
              absorption (left) and emission (right) process
              without change of direction. The wavevectors are
              represented by arrows.
	}
	\label{fig:drawingphasematchingoverviewabs}
\end{figure}

If after absorption the wavevector of the spin wave remains unchanged 
during storage, $\vec{\kappa'}=\vec{\kappa}$ 
and the same control pulse direction $\vec{k}'_\text{c}=\vec{k}_\text{c}$ is used (cf. Fig.~ \ref{fig:drawingphasematchingoverviewabs}), 
clearly the emitted signal pulse retains its original direction
$\vec{k}'_\text{s}=\vec{k}_\text{s}$ as the PDEs from (\ref{eq:pde}) keep applying. 
More generally, the wavevector $\vec{\kappa}'$ stored in the spin wave 
and the wavevector $\vec{k}'_\text{c}$ of the control pulse are 
the only wavevectors that define the direction of re-emission. 
The wavevector of the emitted signal pulse becomes 
\begin{equation}
   	\vec{k}'_\text{s} = \vec{\kappa}'+\vec{k}'_\text{c}. \label{eq:phase_match_em}
\end{equation}
The regarded electric field envelope accordingly changes to
$\mathcal{E}^{\vec{k}'_\text{s}}$ with 
direction of motion
$\vec{e}_{\vec{k}'_\text{s}}$ and accordingly adjusted values in (\ref{eq:def_envelopes}-\ref{eq:pde}). 

The equations (\ref{eq:phase_match_abs}) 
and (\ref{eq:phase_match_em}) are called phase matching conditions, 
as they need to be fulfilled in order to get constructive interference from the different participating atoms. This introduces the spatial extent of the atomic cloud $L$ as parameter that defines how closely the phase matching conditions need to be fulfilled in order to ensure purely constructive interference throughout the cloud. In \ref{subsec:mode_mismatch} we explore these conditions for our regarded system.

For the absorption and emission processes to be efficient,
energy and momentum both need to be conserved.  Energy conservation 
implies that two-wave resonance in 
the atomic $\Lambda$-level system is necessary:
\begin{eqnarray}\label{eq:resonance_cond}
	c |\vec{k}_\text{s}| -c |\vec{k}_\text{c}| =  \omega_{gs} &\quad \text{ for absorption,}\\ \nonumber
	c |\vec{k}'_\text{s}| -c |\vec{k}'_\text{c}|=  \omega_{gs} &\quad \text{ for emission.}
\end{eqnarray}	

\begin{figure}
	\centering
	\includegraphics[width=0.3\linewidth]{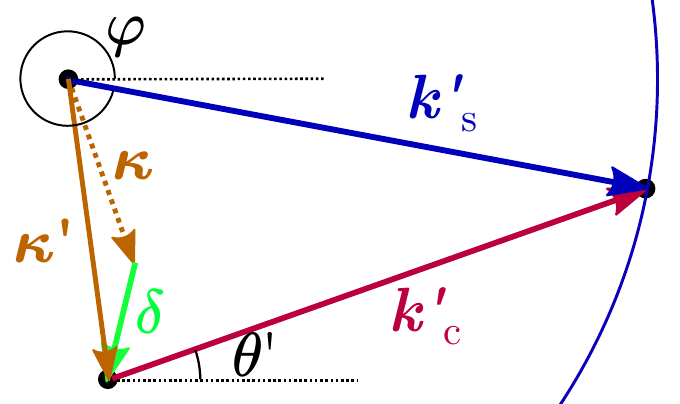}
	\hspace*{.1\linewidth}
    \includegraphics[width=0.3\linewidth]{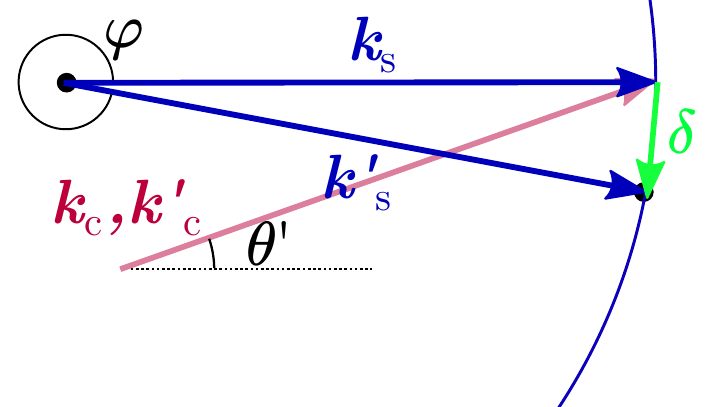}
	\caption{{\em Left:} The phase matching condition for the emission process, 
			when the wavevector stored in the spin wave is changed by $\vec{\delta}$ before emission. 
            The blue segment of a circle marks the 
            wavevectors with
            $|\vec{k}'_\text{s}|=|\vec{k}_\text{s}|$. 
			{\em Right:} Phase matching condition for absorption, manipulation $\vec{\delta}$, and emission for 
			$\vec{k}_\text{c}=\vec{k}'_\text{c}$, i.e.~when the same control beam
			is used for absorption and emission.  
	}
	\label{fig:drawingphasematchingoverviewem_simpleMan}
\end{figure}

These relations allow for the possibility of manipulating 
the emission direction of the signal pulse 
by changing either wavevector on the right hand side of (\ref{eq:phase_match_em}).
Using emission control pulses in different directions was
proposed in \cite{chen_controllably_2016,tordrup_holographic_2008},
but has the disadvantage of 
transferring the problem of controlling the direction of a
light-field from the signal beam to the control beam,
i.e.~one needs active optical elements or different sources
for the control beam.  Here we study the possibility of
changing the wavevector stored in 
the spin wave,
$\vec{\kappa}\rightarrow\vec{\kappa}':=\vec{\kappa}+\vec{\delta}$
(defining $\vec{\delta}$ as ``manipulation''), which can be
done with purely electronic means,
as we will show below.  How this
selects a new direction of the emitted signal pulse is
depicted in
Fig.~\ref{fig:drawingphasematchingoverviewem_simpleMan}:  The
atomic spin wave state starts with the wavevector
$\vec{\kappa}$, is changed by $\vec{\delta}$ to become
$\vec{\kappa}'$; a photon of wavevector $\vec{k}'_\text{c}$
is absorbed, and a photon of wavevector $\vec{k}'_\text{s}$
emitted.  
With this, the direction of emission of the signal pulse
$\vec{k}'_\text{s}$ can deviate from the original direction
$\vec{k}_\text{s}$ even when using the same control beam,
$\vec{k}'_\text{c}=\vec{k}_\text{c}$. The angular change in
direction is denoted by $\varphi$. 

During idealised manipulation, only $\vec{\kappa}$ is changed
to become $\vec{\kappa}'$ without otherwise affecting the spin
wave (see eq.~\ref{eq:manipulation_cond}). 
The exact values of the necessary spin-wave manipulation for inducing
a change in directionality $\varphi$ in the emitted signal pulse are
easily obtained with 
\begin{eqnarray}\label{eq:manip_forw_retrieval}
	\vec{\delta} &= \vec{k}'_\text{s}-\vec{k}_\text{s}
       		= k_\text{s}\ \begin{pmatrix} \cos(\varphi)-1 \\
                             \sin(\varphi)\\ 0 \end{pmatrix}
              ,\quad |\vec{k}_\text{s}| = |\vec{k}'_\text{s}|\,.
	\end{eqnarray}

For small angles $\varphi$, the
increase is linear,  $ \vec{\delta} \approx \varphi
k_\text{s} \vec{e}_y.$ and for large angles it caps at 
$|\vec{\delta}|=2|\vec{k}_\text{s}|$. 
In Sec.~\ref{subsec:mode_mismatch} we study the
decrease in efficiency when
(\ref{eq:manip_forw_retrieval}) is not satisfied exactly.

\subsection{Manipulation via Zeeman shift} \label{subsec:Manip_Zeeman}
The manipulation needed to re-emit the light
into a new direction $\vec{k}'_\text{s}$ can be understood as
the creation of a new spin-wave state that would have resulted from 
signal and control pulses of wavevectors $\vec{k}'_\text{s}$ and
$\vec{k}'_\text{c}$, with unchanged wave numbers
$|\vec{k}'_\text{s}|=|\vec{k}_\text{s}|$,
$|\vec{k}'_\text{c}|=|\vec{k}_\text{c}|$.  
This can be achieved by introducing a
position-dependent phase equivalent 
to a wavevector $\vec{\delta}$: 
\begin{eqnarray}\label{eq:manipulation_cond}
\hat{S}^{\vec{\kappa}'}({ \vec{r},t_2})\ \ket{\psi_{S}^{P} (t_2)} 
		\overset{!}{=}& S({\vec{r},t_1})\ \ket{\psi_{S}^{P} ({t_2})}\\ \nonumber 
\overset{(\ref{eq:def_envelopes})}{\Rightarrow} S({\vec{r},t_2}) 
		\overset{!}{=}& e^{i (\vec{\kappa}'-\vec{\kappa})\cdot \vec{r}} 
				S({\vec{r},t_1})= e^{i \vec{\delta}\cdot \vec{r}} S({\vec{r},t_1}),
\end{eqnarray}
with the manipulation $\vec{\delta}$ leading to emission
angles $\varphi$ as given in
(\ref{eq:manip_forw_retrieval}). 

More generally, arbitrary phases $\phi(\vec{r})$ imprinted on the spin-wave 
such that $S(\vec{r},t_2)=e^{i \phi(\vec{r})} S(\vec{r}, t_1)$ 
can be treated by decomposing the resulting spin-wave into 
separate plane-wave contributions and their envelopes, 
each of which can be described individually by the PDEs 
using the corresponding wavevector $\vec{\kappa}'$. 
The added phases amount to a convolution in $k$-space of the original spin-wave
with the added phase factors $e^{i\phi( \vec{r})}$, as
can be seen from the mathematical relation
$\mathcal{F}\left[ e^{i \phi(\vec{r})} S(\vec{r}, t_1) \right] \propto
\mathcal{F}\left[ e^{i \phi(\vec{r})} \right] 
\star \mathcal{F}\left[ S(\vec{r}, t_1) \right]$, 
where $\mathcal{F}$ denotes the Fourier transform to k-space 
and $\star$ the convolution operator.
While added phases linear in space solely 
shift the wavevector of the spin-wave, 
periodic phase patterns will split the spin-wave into 
several contributions as described 
and demonstrated in \cite{mazelanik_coherent_2019}. 
In our description, $\vec{\kappa}'=\vec{k}'_{\text{s}}-\vec{k}'_{\text{c}}$ 
and $|\vec{k}'_{\text{s}}|=|\vec{k}_{\text{s}}|$ 
must be fulfilled for the derivation of the PDEs to be valid, 
such that other wave-vector contributions to the spin-wave, 
i.e. any mode-mismatch, 
need to be treated as part of the envelope (see \ref{subsec:mode_mismatch}).

A possible way of introducing the
necessary phases is via the Zeeman shift created by 
a magnetic field gradient. For this, we introduce a classical
magnetic field $B(\vec{r},t)$
 of which we assume that it induces an
energy shift in the atomic energy levels that is linear in the
magnetic field. 
In principle, for the regarded cloud of rubidium atoms
this regime can be reached by applying a  
homogeneous magnetic field
$\vec{B}_0\approx {5}\,${kG}~$\vec{e}_z$ that pushes the atomic energy levels
into the Paschen-Back regime, such that
the effect of an additional gradient field leads to
approximately linear responses \cite{sargsyan_hyperfine_2014, steck_rubidium_2015}. 
However, in rubidium this strength of $B_0$ changes the level structure such that our $\Lambda$ scheme is not available. By using a weak magnetic field for the storage and emission processes and ramping up $B_0$ for the duration of the manipulation scheme, the Paschen-Back-regime could still be used to manipulate the spin-wave: As we find in Appendix \ref{sec:adiabaticity}, the adiabaticity condition remains fulfilled for realistic ramp-up speeds, such that the ground states $\ket{g}=\ket{F=1,m_F}$ and $\ket{s}=\ket{F=2,m_F}$ are mapped to the states $\ket{\tilde{g}}=\ket{m_I=m_F+\frac{1}{2}, m_s=-\frac{1}{2}}$ and $\ket{\tilde{s}}=\ket{m_I=m_F-\frac{1}{2}, m_s=\frac{1}{2}}$. In practice, it might be simpler to create the spatially linearly increasing shift of the energy levels in a different way: with the use of a spatially non-linearly increasing magnetic field that accounts for the non-linear response of the atoms, the necessary effect can be induced without need for a fully linear response to additional magnetic fields as assumed here. This avoids the need to change $B_0$ before and after the spin wave manipulation.
For an order-of-magnitude estimation, we nonetheless regard the linear regime with the Hamiltonian
\begin{equation}
\hat{H}_\text{B} = -\sum_{i} B(\vec{r}_i,t) 
		\left( \mu_{g}\hat{\sigma}^{i}_{gg} + \mu_{e}\hat{\sigma}^{i}_{ee} 
				+\mu_{s}\hat{\sigma}^{i}_{ss} \right),
\end{equation}
with $\mu_{x}$ being the respective magnetic moment
corresponding to the atomic states $x\in\{ g, e, s \}$. 
The induced energy shifts lead to a changed time evolution
during the storage time, which is solved by
\begin{equation}
S({\tilde{\vec{r}}, \tilde{t}_2}) = e^{i
         	\phi_\text{tot}({\vec{\tilde r 
         	}})} S({\tilde{\vec{r}}, \tilde{t}_1}), \label{eq:Srt2}
\end{equation}
where $\tilde{t}_1$ and $\tilde{t}_2$ are the initial and final 
regarded moments in rescaled time and 
\begin{equation}\label{eq:manip_accum_phase}
\phi_\text{tot}({\tilde{\vec{r}}}):= (\mu_{g}-\mu_{s})/(\gamma\hbar)\  
		\int_{\tilde{t}_1}^{\tilde{t}_2}d \tilde{t}\  B({\tilde{\vec{r}},\tilde{t}})
\end{equation} 
is the locally accumulated phase in the spin wave 
due to the magnetic field. Any global phase can be ignored. Thus, the necessary property of the $g$ and $s$ levels for our Zeeman manipulation to be applicable is that the two states differ in their reaction to magnetic fields, i.e. $\mu_g\neq \mu_s$ in our notation. 
This condition is indeed fulfilled for alkali atoms with hyperfine-split ground states and sufficiently weak magnetic fields. For schemes using atomic clock states with a suitably chosen value of $B_0$ to minimize the susceptibility of the spin-wave to stray magnetic fields (i.e. $\mu_g=\mu_s$), changing the strength of $B_0$ for the duration of the manipulation can still allow for the Zeeman manipulation scheme to be applied, while of course the spin-wave will be susceptible to stray magnetic fields for that duration.
Inserting (\ref{eq:Srt2}) into
    (\ref{eq:manipulation_cond}) gives 
\begin{eqnarray}
      \phi_\text{tot}({\tilde{\vec{r}}}) =& \vec{\delta}\cdot 
                                              \tilde{\vec{r}} L+const.\\
      \nonumber 
       \Leftrightarrow & \int_{t_1}^{t_2}d t\  B(\vec{r},t) 
          		= \frac{\hbar \vec{\delta}\cdot \vec{r} }{\mu_{g} - \mu_{s}}+const.\,.
\end{eqnarray}
For simplicity, we regard the time needed for manipulation using a fixed field gradient.
The direction of the needed
gradient of the magnetic field-amplitude $B$ is given by
(\ref{eq:manip_forw_retrieval}) and we denote 
the contribution of $\vec{r}$ parallel to $\vec{\delta}$ 
with $r_{\parallel\vec{\delta}}$.  
With a field $B({\vec{r}})=B_0+{50}\, \frac{\text{G}}{\text{cm}}\cdot 
r_{\parallel\vec{\delta}}$,
duration $T$ and coupling corresponding to an electronic spin transition \cite{steck_rubidium_2015}
\begin{eqnarray}\nonumber
(\mu_{g} - \mu_{s})/\hbar \approx 2\mu_\text{Bohr}/\hbar \approx {17.6}\,\text{rad/}\mu\text{s/G}.
\end{eqnarray}
For rubidium, this approximate value is reached both for weak magnetic fields and in the regarded Paschen-Back-regime.
	
This gives
\begin{equation}\label{eq:T}
T =\dfrac{\hbar}{\mu_g-\mu_s}\cdot\dfrac{|\vec{\delta}|}{50\,\text{G/cm}} =  \frac{|\vec{\delta}|}{88/\text{mm}}\,\mu\text{s},
\end{equation}
which leads to necessary manipulation times of the order of
$T\approx 10^{-4}$\,s to achieve arbitrary angles $\varphi$. 
A finite speed in turning on and off the field gradient 
will increase the necessary time correspondingly.

The decoherence time scale 
from thermal motion of freely moving atoms at different temperatures and the corresponding limitations to the reachable deflection angles 
are discussed at the end of \ref{subsec:mode_mismatch}. 
We find that deflection angles $\varphi\approx 20\,$mrad remain viable in cold atomic clouds, 
but arbitrary deflection angles will likely require a different system. E.g. dopants in a solid body can act as a suitable
atomic ensemble \cite{wang_three-channel_2009, lvovsky_optical_2009} 
where diffusion does not occur. 
 
In order to achieve arbitrary deflection angles $\varphi$ on the order of $\mu$s, 
correspondingly the rather large field gradient of $50\,$G/cm 
has to be created on a similar time scale. 
Using Maxwell coils \cite{hidalgo-tobon_theory_2010}, 
a rise time of $5\,\mu$s can be achieved with $63$ turns, 
a coil radius of $1\,$cm and a maximum current of $1\,$A, 
while using a current source delivering $<40\,$V. 
With the focus on small deflection angles, 
a smaller maximum gradient of $7\,$G/cm can be chosen. 
Using the same current source this allows for a much faster rise time 
of $0.1\,\mu$s which allows deflection angles of up to $\sim 0.2\,$mrad 
at thermal velocities of room-temperature vapors.
A more detailed description of the coil parameters can be found in appendix \ref{seq:coil_properties}. In appendix \ref{sec:adiabaticity} we confirm that adiabaticity remains fulfilled in the regarded parameter regime such that, apart from the intended phases, the state of the system is not significantly affected by the field gradient.

\section{Numerical Results}
\label{sec:Num_results}

\begin{figure}
	\centering
	\includegraphics[width=0.7\linewidth]{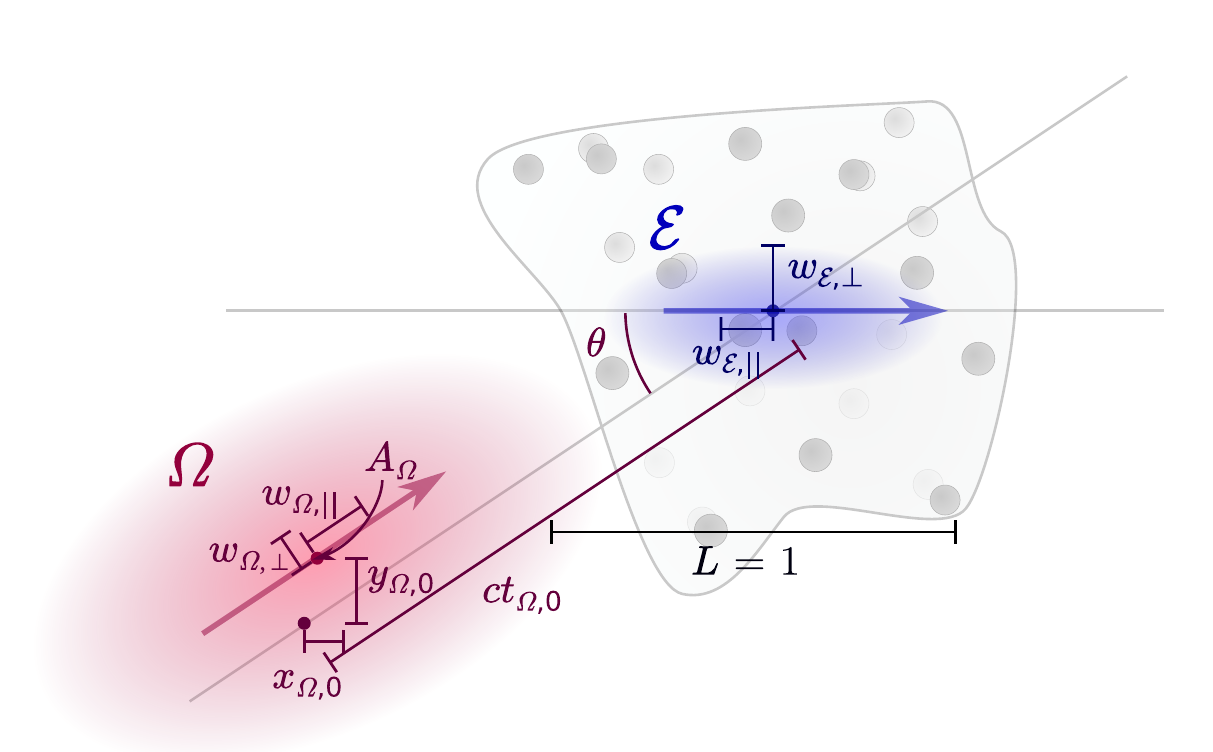}
	\caption{Relevant parameters that define the incoming
		signal and control pulses. 
		\label{fig:drawing_param_abs}}
\end{figure}

In the following we provide results from solving
(\ref{eq:final_pde}) numerically and optimizing the efficiency
with which pulses can be stored and re-emitted in different
directions. 
For simplicity, we restrict the incoming signal 
and control pulse to Gaussian shape 
with widths $w_{\mathcal{E},\parallel}$ 
and $w_{\Omega,\parallel}$ parallel to the respective direction of propagation,  
and the corresponding orthogonal beam widths 
$w_{\mathcal{E},\perp}$ and $w_{\Omega,\perp}$. 
The signal pulse is chosen to propagate along the $x$-axis, 
reaching the cloud's center at $t=0$. 
The control pulse propagates at an angle $\theta$ relative to 
the signal pulse and its timing and position are parametrized such that 
at time $t_{\Omega,0}$ the position of its peak 
is $(x_{\Omega,0},y_{\Omega,0})$ in the xy-plane. 
$A_{\Omega}$ denotes the amplitude of $\Omega$. 
The parameters are drawn in Fig.~\ref{fig:drawing_param_abs}. 
The results of \cite{gorshkov_photon_2007-1} and
\cite{gorshkov_photon_2007-2} allow one to get estimates of the  
{scaling of the reachable efficiency with optical depth.} 
The achievable efficiencies are in general upper bounded by
efficiencies that can be reached with the help of a cavity
that restricts the electric field to a single relevant spatial mode
\cite{gorshkov_photon_2007-1}, 
\begin{equation}\label{eq:eta_scaling_cavity}
	\eta^\text{max}_\text{cavity} \le \left( \eta^\text{max}_\text{abs, cavity} \right)^2 = \left( 1-\frac{1}{1+d'} \right)^2\,,
\end{equation}
which hence provides an important benchmark.

For high optical depths the reachable efficiency in free space can be approximated by
\begin{equation}\label{eq:eta_scaling}
	\eta^\text{max} \le \left( \eta^\text{max}_\text{abs} \right)^2 \overset{d\rightarrow\infty}{\sim} \left( 1-\frac{2.9}{d'} \right)^2.
\end{equation}
We choose
\begin{equation}
      \label{eq:ref}
	\eta^\text{ref} =\left( 1-\frac{1}{1+d'/2.9} \right)^2
\end{equation}
as reference for our results 
as it has an optical depth-dependence similar to (\ref{eq:eta_scaling_cavity}) 
and becomes an approximate upper bound for $d\rightarrow \infty$. 
As the chosen numerical method matches the discretised coordinates $\tilde{x}$ 
and $\tilde{c}\tilde{t}$ in order to achieve 
a simple propagation of $\mathcal{E}$ in (\ref{eq:final_pde}), 
the length of the regarded incoming signal pulses is limited 
due to computational constraints. 
Thus, the regarded signal
pulses are of high bandwidth $\Delta \omega_\text{s} \gg \gamma$ with 
\begin{equation}\label{fig:effective_pulse_bandwidth}
\frac{\Delta \omega_\text{s}}{\gamma} = 
		\frac{\tilde{c}}{\tilde{w}_{\mathcal{E},\parallel}}.
\end{equation}
We expect high values of
$\tilde{c}/\tilde{w}_{\mathcal{E},\parallel}$ to negatively
affect the reachable efficiency as increasingly short pulses 
make higher optical depths necessary in order to reach 
optimal efficiency \cite{gorshkov_photon_2007-2}. 

We use parameters corresponding to a uniform, 
spherical cloud of $^{87}{\text{Rb}}$ with volume $V=L^3=({10}$\,{mm}$)^3$ and $\tilde{c}=850$.
Unless explicitly stated otherwise, parameter values for the signal
pulse are $\tilde{\Delta}=0.0, \tilde{w}_{\mathcal{E},\parallel}=100,
\tilde{w}_{\mathcal{E},\perp}=0.2$, while the control parameters 
(i.e. width $w_{\Omega,\perp}$, length $w_{\Omega,\parallel}$, 
amplitude $A_{\Omega}$, timing $t_{\Omega,0}$ and displacement $x_{\Omega,0}$) are
optimized to give high efficiencies. This corresponds to a high
frequency bandwidth of the signal pulse 
$\Delta \omega_\text{s}  \approx {0.3}$\,{GHz}, which makes the
parameter regime comparable to the Autler-Townes storage scheme in
\cite{saglamyurek_coherent_2018} except that control pulses with
similar dimensions as the signal pulse are used. Note that although the limitations for higher signal bandwidths are not visible in the PDEs, for rubidium there are limitations as higher signal bandwidths will require changes to the $\Lambda$-system as the hyperfine coupling is no longer stronger than the necessary coupling to the light fields and additionally there arises significant overlap in the spectrum of the control and signal field.
The choice of $\Delta=0$ is made for numerical simplicity.  

Before regarding the full process consisting of absorption, storage
and reprogramming of direction, and emission, we study the
absorption processes separately, in particular with respect to the
achievable absorption efficiencies as function 
of the angle $\theta$ between signal and
control beam.  

\subsection{The absorption process}
\label{subsec:numerical_res_absorption}
\begin{figure}
	\centering
	(a) 
	\includegraphics[width=0.45\linewidth]{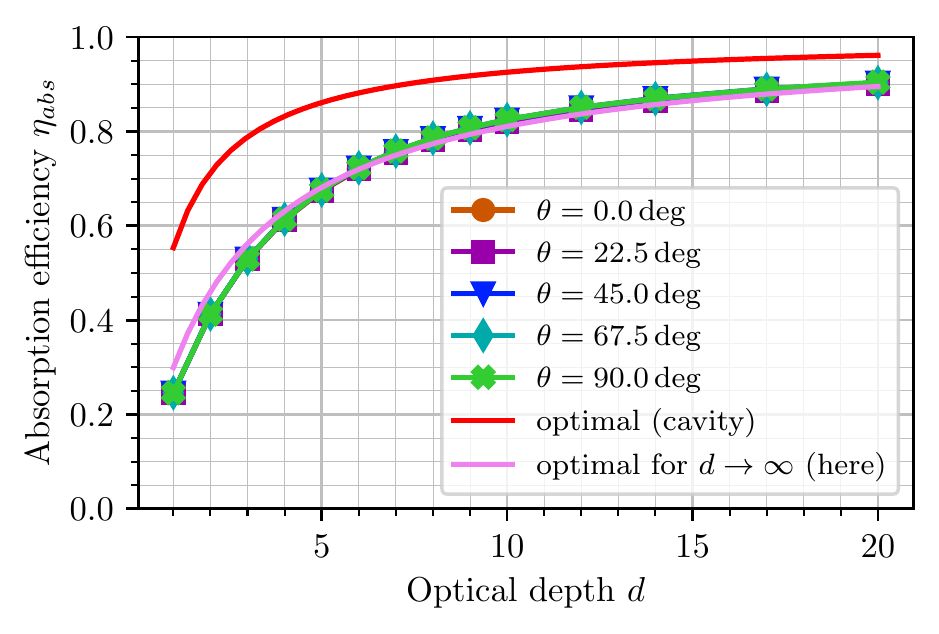}
	(b) \includegraphics[width=0.45\linewidth]{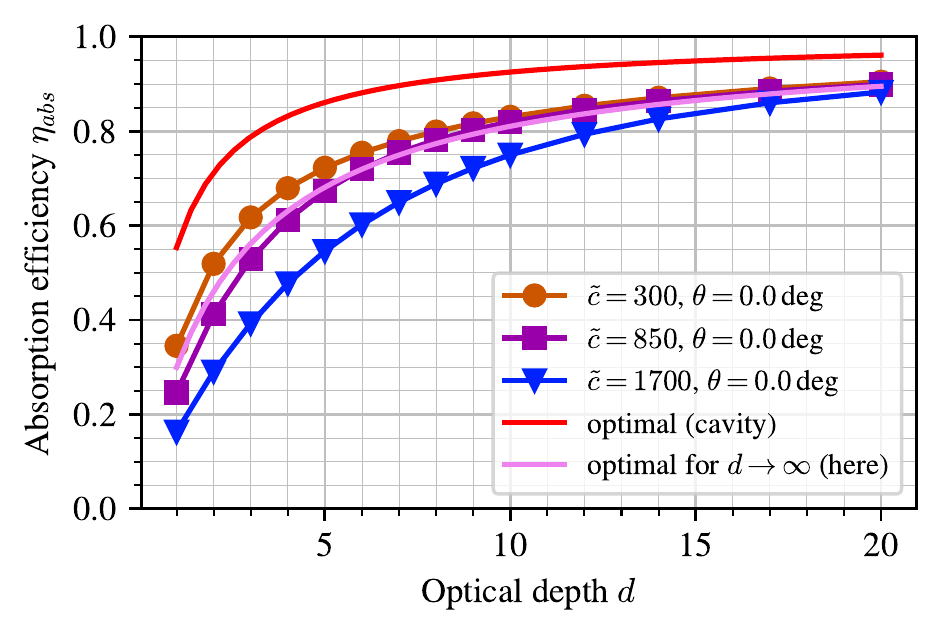}
	(c)\includegraphics[width=0.45\linewidth]{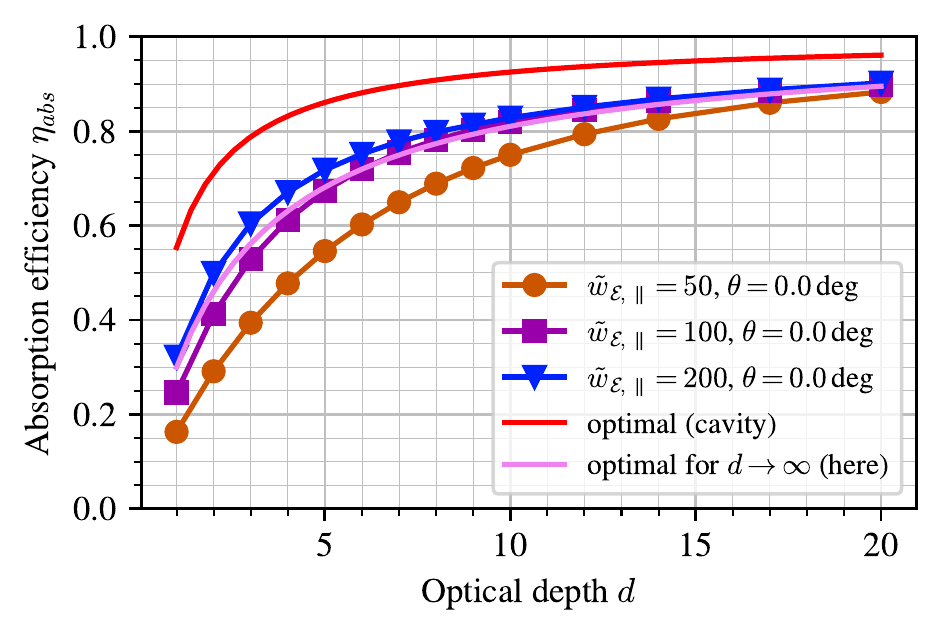}
	(d)\includegraphics[width=.45\linewidth]{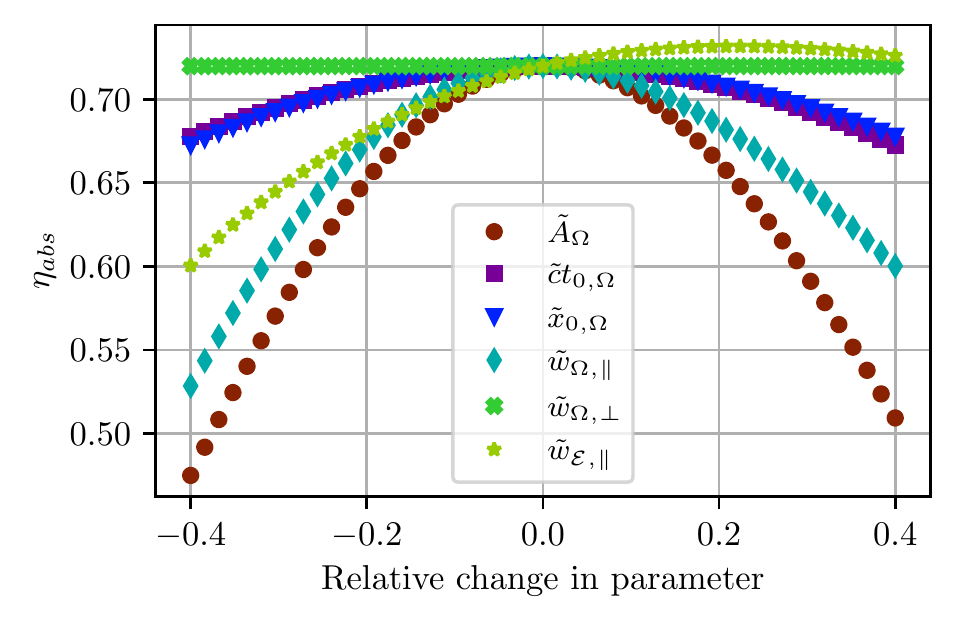}	
	\caption{Maximum absorption efficiencies as function
              of different parameters.  (a):
              $\eta_{abs}$ achieved for various values of optical
              depth $d$ and signal-to-control angle $\theta$ for
              Gaussian pulses. The achievable efficiency (in the
              cavity case) is shown in red (dark gray), the free-space reference curve is plotted in pink (light gray).
              (b): influence of cloud size on
              reachable absorption efficiencies.
              (c): influence of signal pulse length on
              reachable absorption efficiencies.	
              (d): robustness of efficiency to single
              parameter variation using values from (a) as reference, with $d=6$ and $\theta=0$.} 
	\label{fig:sys3effabsall}
\end{figure}

For testing the achievable storage efficiencies, a simple
optimisation of control pulse parameters for varying values of
$d$ and $\theta$ was done. The results are given in
Fig.\ref{fig:sys3effabsall}.  Fig.\ref{fig:sys3effabsall}(a)
shows that 
efficiencies comparable to our reference curve from (\ref{eq:eta_scaling}) are
already reached for $d\approx 5$, while the angle between signal 
and control pulse $\theta$ does not affect the reached efficiency. For $d=20$ an absorption
efficiency of about 90\% should be achievable for
$\tilde{w}_{\cal{E},\parallel}=100$, $\tilde{c}=850$. 
In Fig.~\ref{fig:sys3effabsall}(b) the reached efficiencies for 
different values for $\tilde{c}=c/(\gamma L)$ are shown, 
which corresponds to altering the size of the atomic cloud 
and Fig.~\ref{fig:sys3effabsall}(c) shows the corresponding results 
for different signal pulse lengths $\tilde{w}_{\mathcal{E},\parallel}$ 
and thus band widths.

Together, Fig.~\ref{fig:sys3effabsall}(b) and (c) confirm 
that high values of ${\tilde{c}}/{\tilde{w}_{\mathcal{E},\parallel}}$
make higher optical depths necessary in order to reach high efficiencies.
Fig.~\ref{fig:sys3effabsall}(d) 
shows the very smooth dependence of the resulting storage efficiency 
on single-parameter-variation. 
As reference parameters, the optimized values corresponding to  
Fig.~\ref{fig:sys3effabsall}(a) at the point $\theta=0,$ $d=6$ were used.

\subsection{Absorption, storage, and re-emission}
\label{subsec:numerical_res_full_storage}
\begin{figure}
\centering
(a)\includegraphics[width=0.45\linewidth]{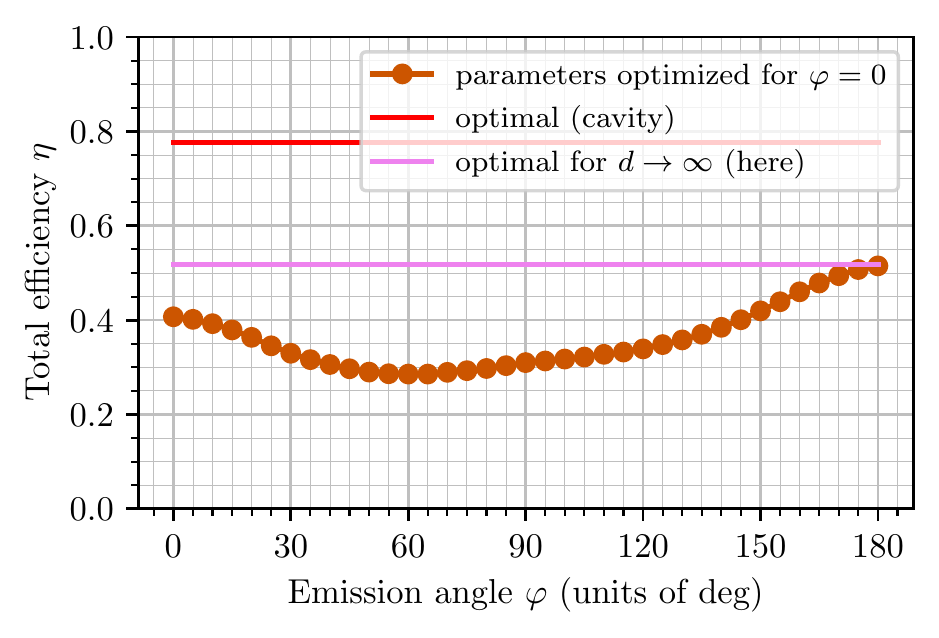}
(b)\includegraphics[width=0.45\linewidth]{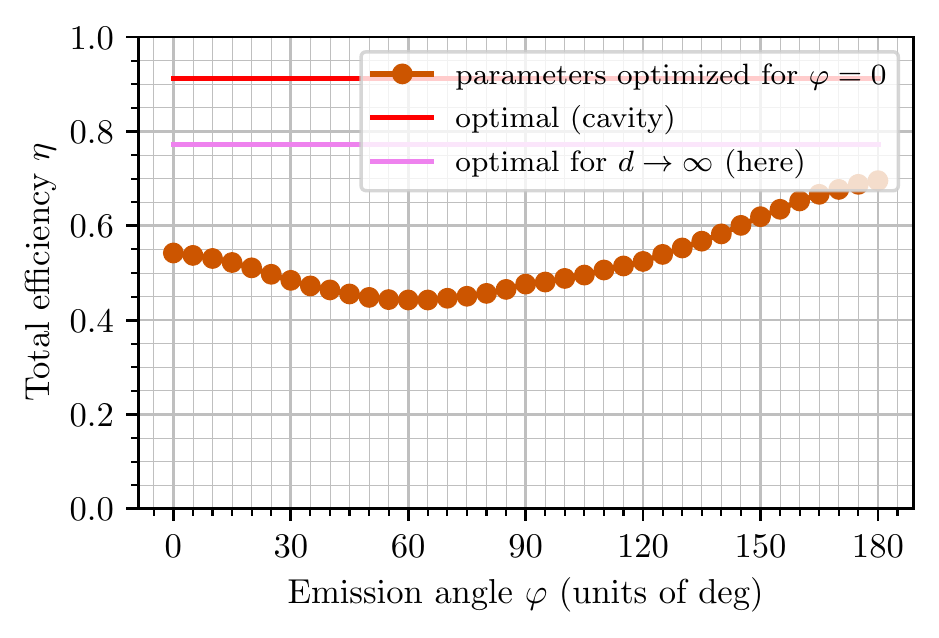}
(c)\includegraphics[width=0.45\linewidth]{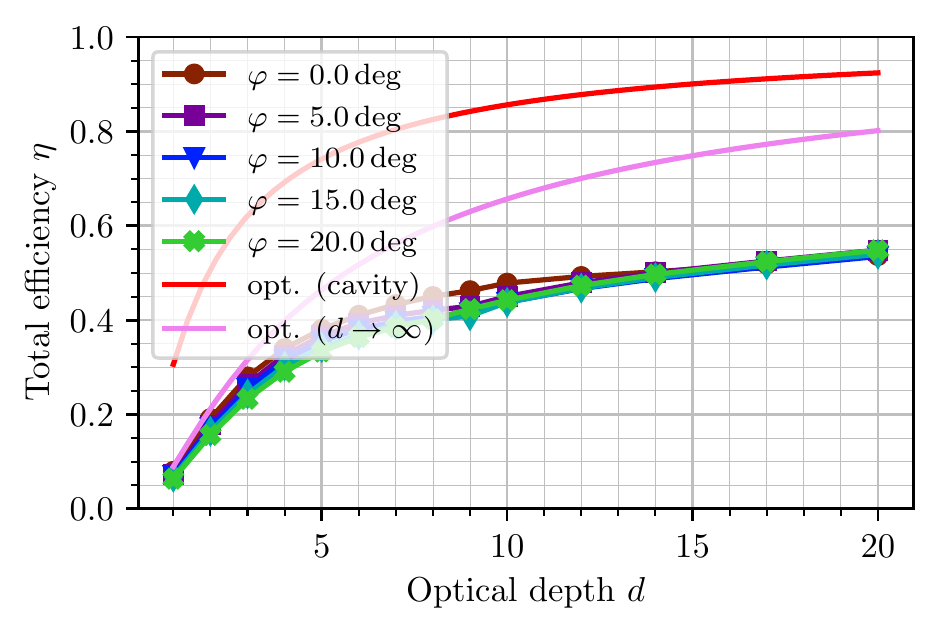}
(d)\includegraphics[width=0.45\linewidth]{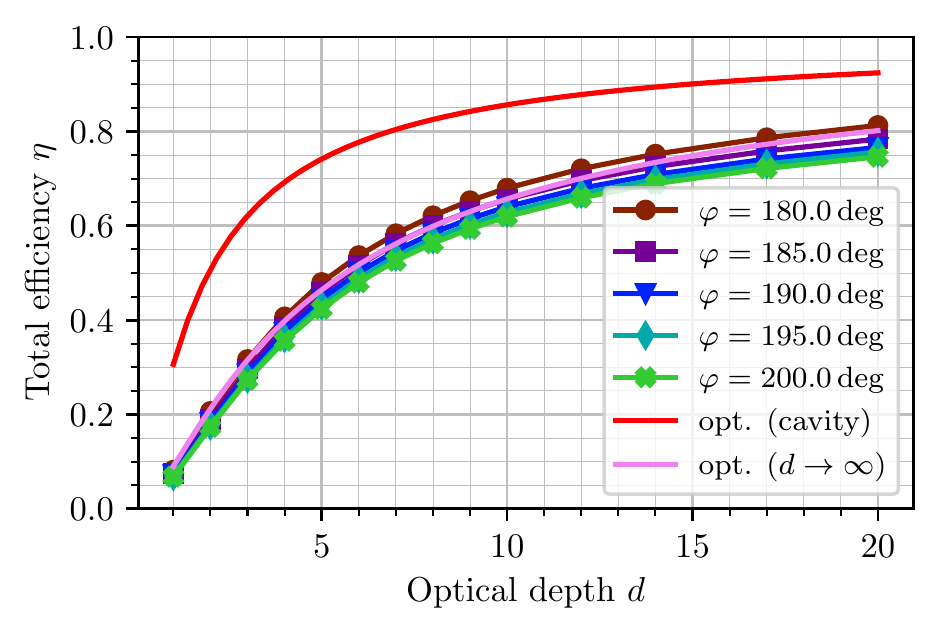}
\caption{(a,~b): Total efficiencies achieved for
 		different re-emission angles $\varphi$ when  using the parameters
  		optimized for $\varphi=0$ and $\theta=0$. (a) uses $d=6$, while (b) uses $d=17$. 
		(c,~d): Achieved efficiencies with parameters optimized for
		each $\varphi$ separately for angles 
		close to $\varphi=0$ for (a) and close to $\varphi=\pi$ for (b).}
\label{fig:optemw100c850d6-17th0}
\end{figure}

\begin{figure}
\centering
(a)\includegraphics[width=.45\linewidth]{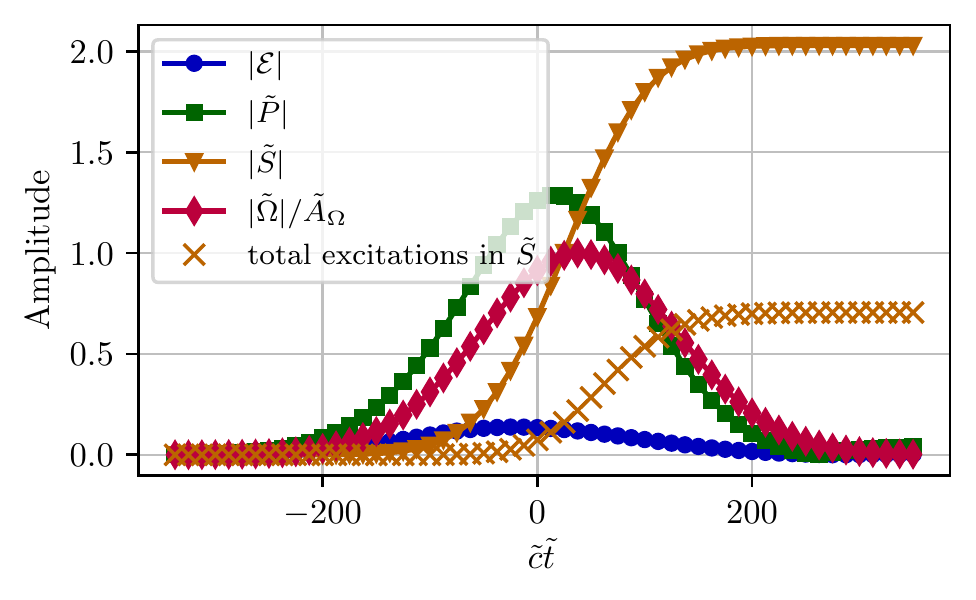}
(b)\includegraphics[width=.45\linewidth]{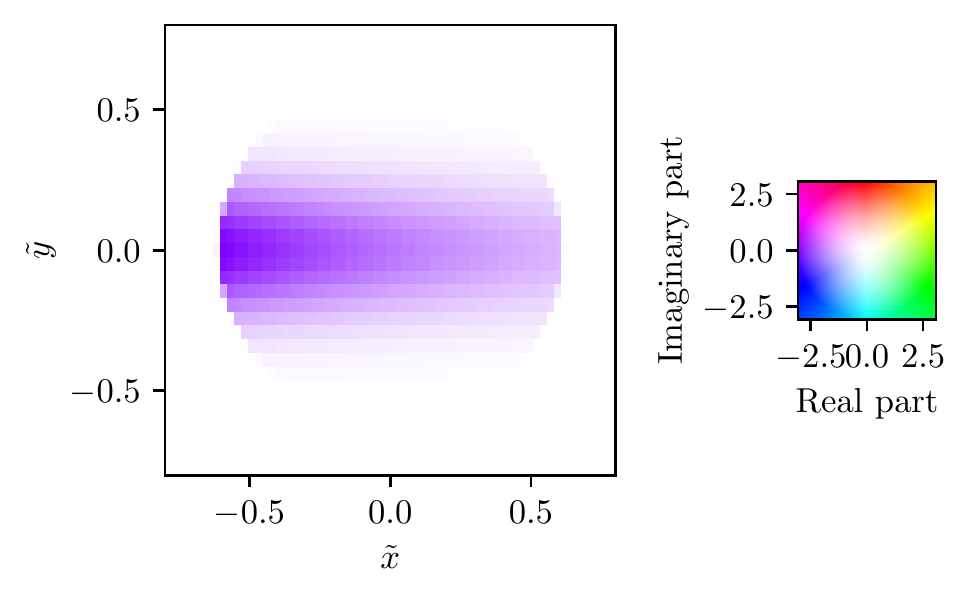}\\
(c)\includegraphics[width=.45\linewidth]{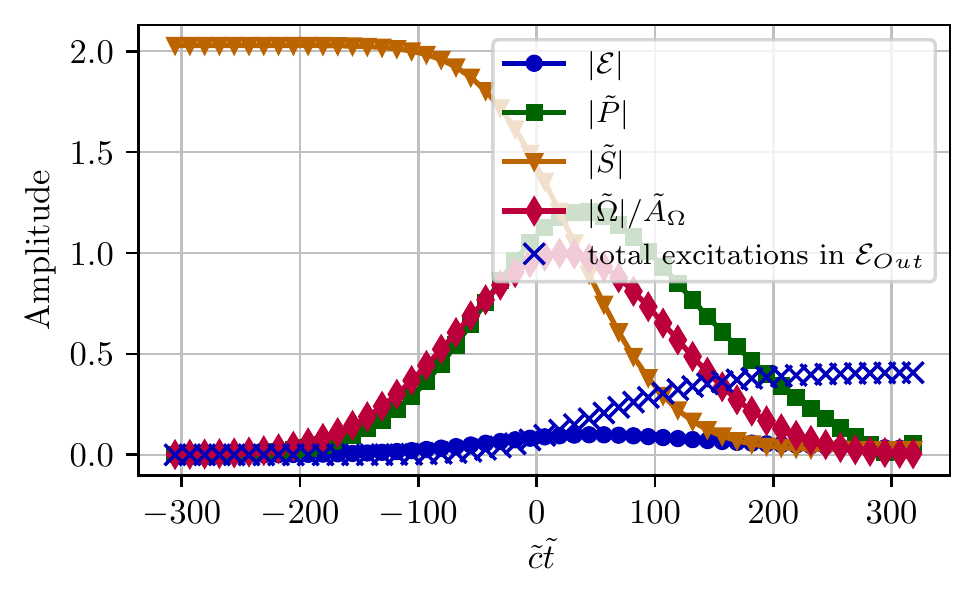}
(d)\includegraphics[width=.45\linewidth]{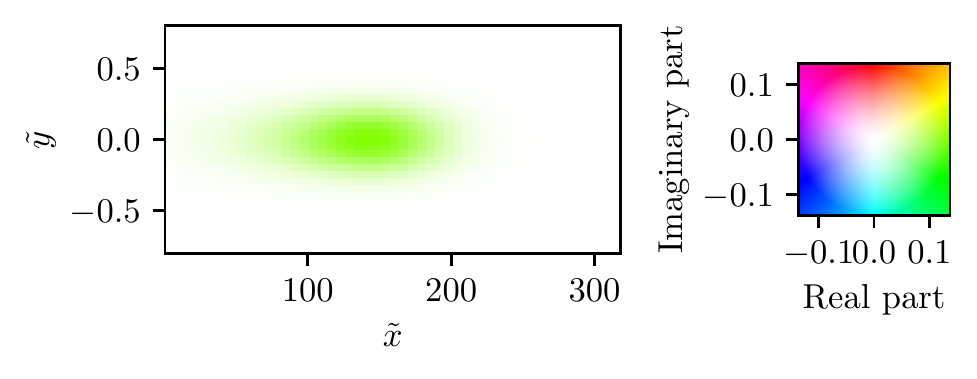}
\caption{Field amplitudes for the full storage process for 
$d=6$, $\varphi=0$ and $\theta=0$.
(a): Amplitude over time of the variables at the center of the cloud for
absorption.  (c): Same for emission. (b): Resulting spin wave after absorption. (d):
Outgoing field envelope after the emission process. 
} 
\label{tab:param_em_w100_d6_th0}
\end{figure}

We now consider the full process of absorption, storage, and
re-emission. For calculating the total efficiency $\eta$, 
the number of re-emitted excitations up to a certain time after 
arrival of the emission control pulse was used, 
such that an altered shape of the re-emitted pulse 
does not affect the calculated efficiency. 
Fig.~\ref{fig:optemw100c850d6-17th0} shows the achieved total efficiencies 
as function of $\varphi$ when using control pulses optimized for
$\varphi=0$. For an optical depth $d=17$,
efficiencies varying between about 45\% and 70\% can be realized, with
a maximum efficiency for backward re-emission ($\varphi=180^\circ$).
Optimizing the parameters separately for each angle can still increase
the efficiencies, in particular for high re-emission angles close to backward emission, 
as can be seen when comparing Fig.~\ref{fig:optemw100c850d6-17th0}(b) and (d). 
As the shape of the signal pulse orthogonal to its direction of propagation 
is preserved during absorption, departing from Gaussian beam profiles 
can improve the achievable emission and thus total efficiencies for intermediate values 
of $\varphi$ as the 
pulse shape originally in the direction orthogonal to propagation
now contributes to the longitudinal shape of the spin 
wave when taking the new direction as reference. 
The amplitudes of the fields $\cal E$, $P$, $S$, and $
\Omega$ as function of space and time that result from the
optimization of the overall efficiency are shown for a typical example ($d=6$ and
$\varphi=0$ from Fig.~\ref{fig:sys3effabsall}(a)) 
in Fig.~\ref{tab:param_em_w100_d6_th0}, both for the absorption and
emission part.  One sees directly how the photon is transferred to a
spin-wave excitation during absorption, whereas the excited state
$\ket{e}$ is only excited very slightly and only for relatively short
time. In emission, the process is inverted, and the excitation of the
spin wave re-converted into an optical excitation.  We also see that
the spin wave envelope $S$ has essentially the same phase over the cross section
of the sample as in the center of the sample, and the same is true for 
the signal pulse that is re-emitted. 

\subsection{Imperfections}
\label{subsec:mode_mismatch}
\begin{figure}
	\centering
	(a)\includegraphics[width=0.45\linewidth]{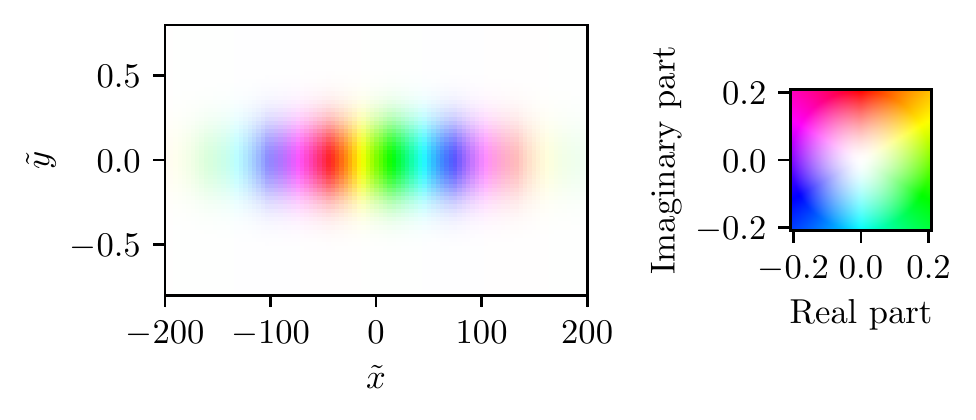}
	(b)\includegraphics[width=0.45\linewidth]{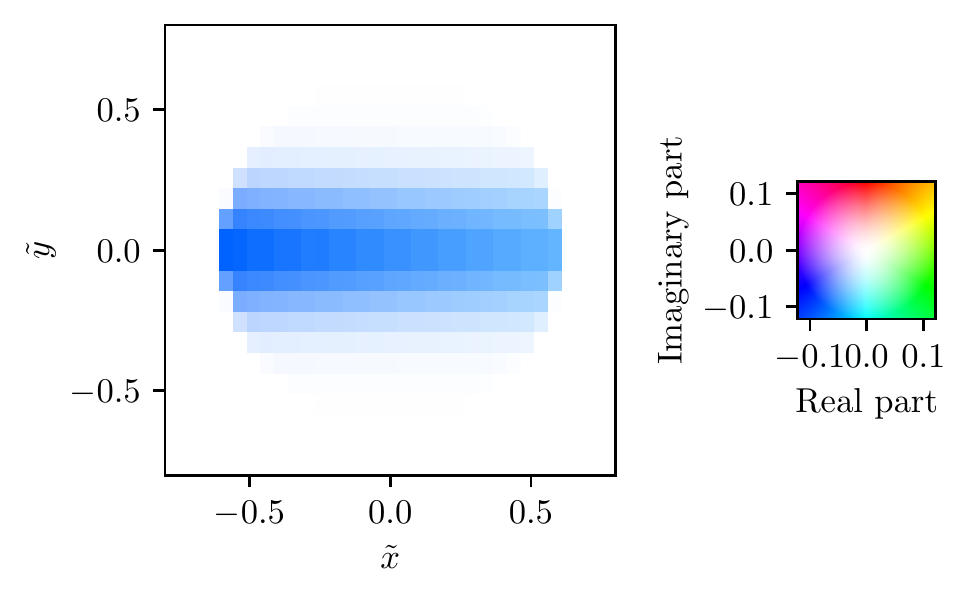}
	\caption{The incoming signal field envelope (a) 
			and resulting spin wave (b) for parameters 
			from Fig.~\ref{fig:sys3effabsall}(a) with $\theta=0$, $d=6$ 
			and mode mismatch  $k_\text{mis}= 30\ \gamma/c$. While the mode mismatch leads to a large range of phases over the extent of the incoming signal pulse, the resulting spin wave has almost constant phase. (colors online).}
	\label{fig:modemismatch_E_In_S}
\end{figure}

\begin{figure}
	\centering
	\includegraphics[width=0.7\linewidth]{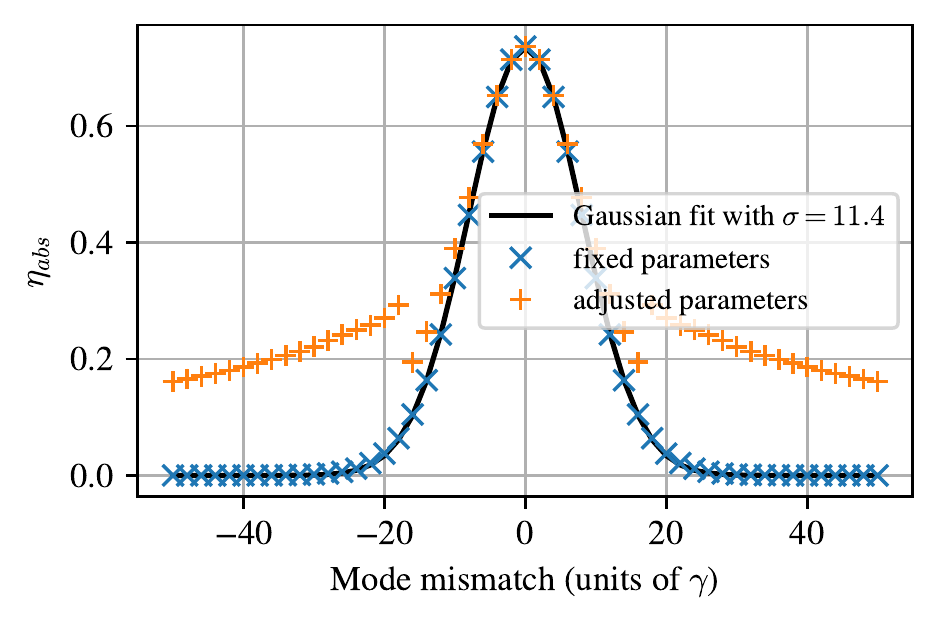}
	\caption{The dependence of the resulting absorption efficiency 
			of the signal field mode mismatch. 
			The corresponding frequency shift is measured 
			in multiples of the spontaneous emission rate $\gamma$.}
	\label{fig:modemismatch}
\end{figure}

For all previous considerations, exact two-wave resonance was assumed, namely
\begin{equation}
	c|\vec{k}_\text{s}|-c|\vec{k}_\text{c}| = \omega_{ge} - \omega_{se}.
\end{equation}
Now we examine the influence of a slightly detuned signal field
with a changed frequency $c|\vec{k}_\text{s}|=
\omega_{ge}-\Delta+ck_\text{mis} $, where $k_\text{mis}$ is
the mode mismatch. The control field frequency remains
$c|\vec{k}_\text{c}|= \omega_{se}-\Delta$. 
A visualisation of a mismatched incoming probe pulse and the resulting spin wave is shown in Fig.~\ref{fig:modemismatch_E_In_S}. With the
spontaneous emission rate of the excited state $\gamma/c$ as reference
and assuming all other parameters as constant, we
find Gaussian suppression of the absorption efficiency 
(see Fig.~\ref{fig:modemismatch}),
\begin{equation}
	\eta_\text{abs}({k_\text{mis}}) \approx \eta_\text{abs}({0})\  
			\exp(-\frac{k_\text{mis}^2}{(11.4\ \gamma/c)^2}).
\end{equation}
Adjusted control parameters can largely compensate 
the exponential suppression 
of efficiency in the regarded range of mode mismatch 
(see the orange pluses in
Fig.~\ref{fig:modemismatch}).

\begin{figure}
\centering
\includegraphics[width=0.7\linewidth]{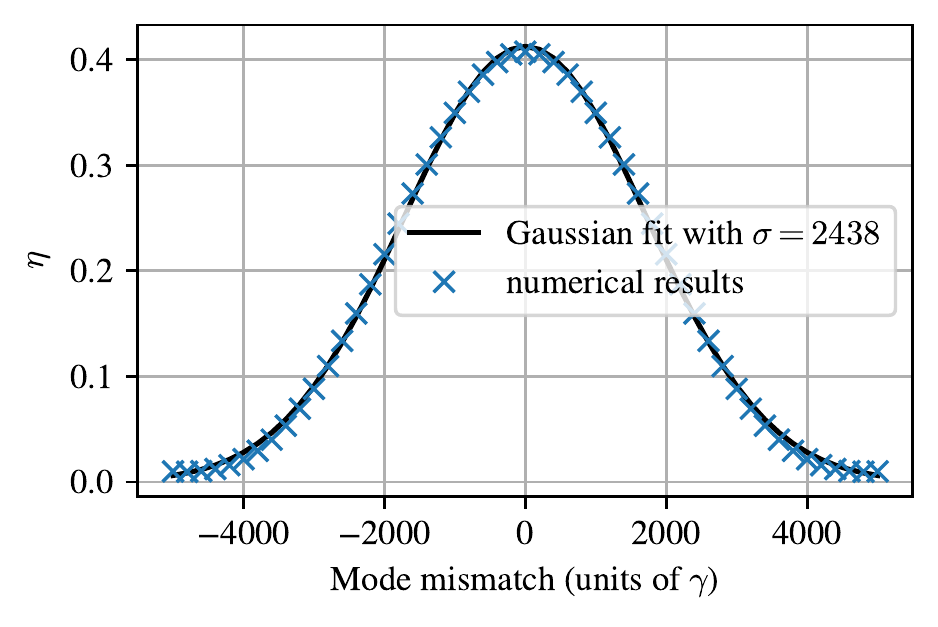}
\caption{The dependence of the resulting total storage efficiency on
		  the spin-wave phase error, e.g. stemming from non-optimal
		  manipulation. The corresponding frequency shift is measured in
		  multiples of the spontaneous emission rate $\gamma$.} 
\label{fig:modemismatch_em}
\end{figure}

\begin{figure}
	\centering
	\includegraphics[width=0.8\linewidth]{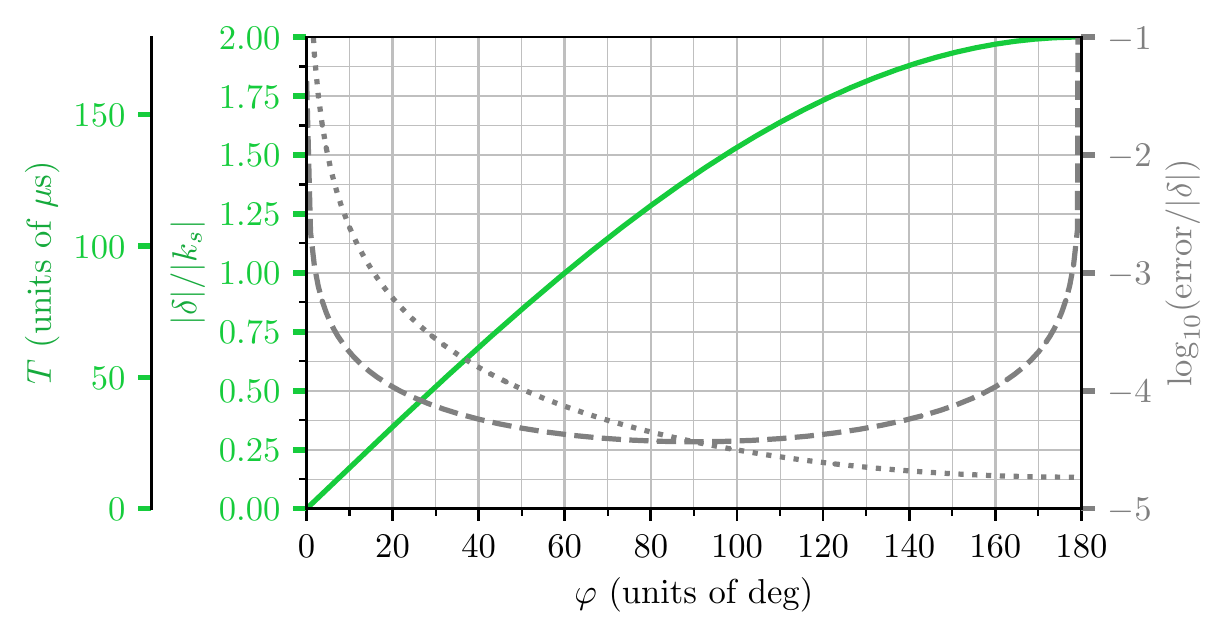}
	\caption{wave number and the corresponding manipulation time 
		(using parameters from \ref{subsec:Manip_Zeeman}) 
		necessary for changing the direction of the signal pulse
		by the angle $\varphi$ (green continuous line).   
		The gray lines indicate the needed accuracy 
		(relative error in $\delta$ that leads to one standard deviation 
		in the Gaussian (\ref{eq:Gauss205}) 
		(dotted line for the error $\parallel \vec{\delta}$, 
		dashed line for the error $\perp \vec{\delta}$)).} 
	\label{fig:manipulationovervarphi}
\end{figure}

When re-emitting the excitation
stored in the spin wave, there might also be a mode mismatch
from a mismatch remaining from the absorption process 
or through non-optimal manipulation $\vec{\delta}$. 
If a mode mismatch is present, 
the momentum and energy conservation conditions 
from (\ref{eq:phase_match_em},~\ref{eq:resonance_cond}) cannot be fulfilled 
and the efficiency diminishes as destructive interference occurs. 
Fig.~\ref{fig:modemismatch_em} shows the decrease of total efficiency 
if a mode mismatch $k_\text{mis}$ is introduced to the stored spin wave according to
\begin{equation}\label{eq:mode_mismatch_S}
	\tilde{S}(\vec{r}) \rightarrow  
			e^{i (k_\text{mis} \vec{e}_{\vec{k}'_\text{s}})\cdot \vec{r}} \tilde{S}(\vec{r}).
\end{equation}
Not changing any other parameters 
(and using the parameters from Fig.~\ref{tab:param_em_w100_d6_th0}), 
the resulting efficiency for forward retrieval shows 
an approximately Gaussian dependence on $k_\text{mis}$,
\begin{equation}
	\eta({k_\text{mis}})\approx  \eta({0})\ 
			\exp(- \frac{k_\text{mis}^2}{(2.9/L)^2} ).\label{eq:Gauss205}
\end{equation} 
As $|\vec{k}_\text{s}|\times L \approx 10^5$ with the parameters used, 
the phase matching condition needs to be fulfilled 
with relatively high precision (see Fig.~\ref{fig:manipulationovervarphi}).
Similarly to the absorption process, 
we expect that the reduction in achievable emission
efficiency can be alleviated by adjusting the control parameters.

\begin{figure}
	\centering
	\includegraphics[width=0.6\linewidth]{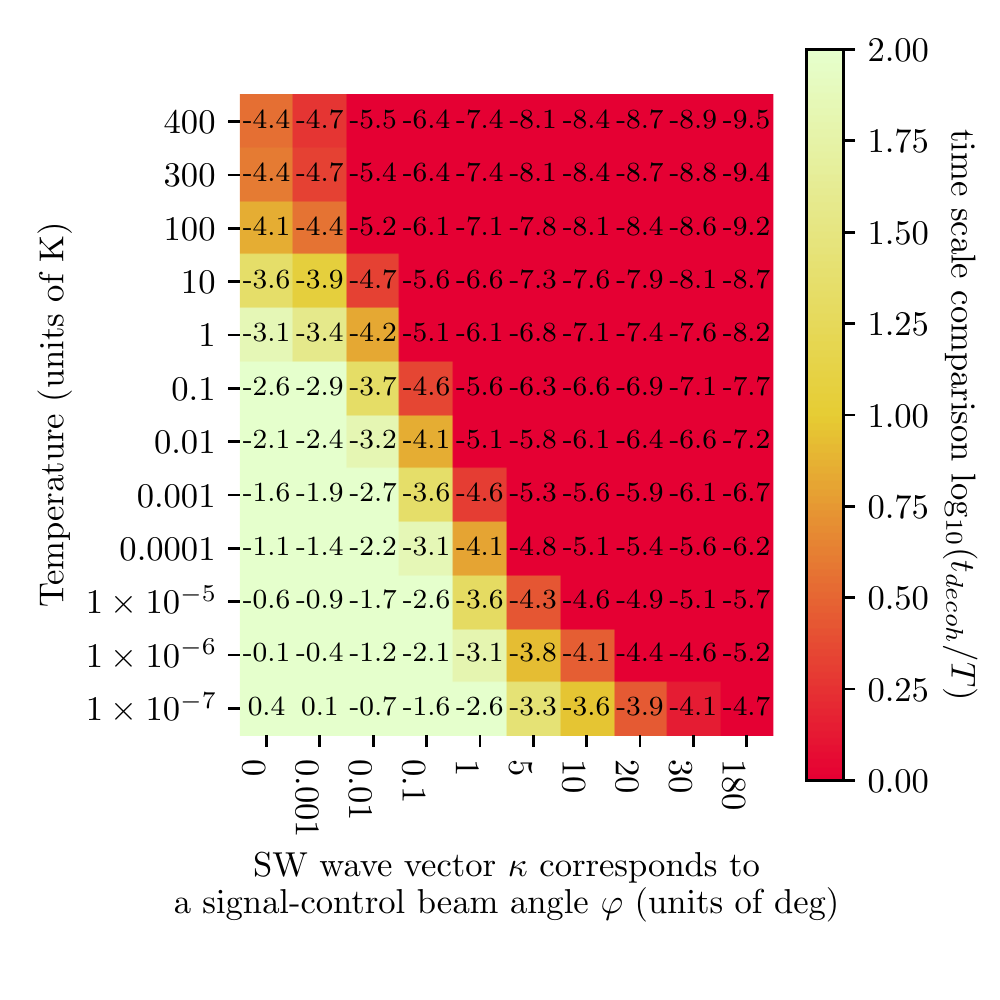}
	\caption{The values printed inside the heat plot are the time scale 
		($\log_{10}(t_{\text{decoh}}/\mathrm{s})$) for spin wave decoherence 
		due to ballistic thermal motion scrambling the phases. 
		The values are given for different temperatures of the Rb atoms 
		as well as wave numbers carried by the spin wave. 
		The color coding indicates whether this motional decoherence leaves sufficient time 
		for our proposed Zeeman manipulation method to imprint or remove wavevectors 
		of the corresponding magnitude using a coil as described in \ref{subsec:Manip_Zeeman}.}
	\label{fig:t_motion_decoh}
\end{figure}

As the spin wave contains phases corresponding to 
the wavevector $\vec{\kappa}$ (see (\ref{eq:field_op_eigenstate})), 
atomic motion scrambling the phases  \cite{zhao_long-lived_2009, saglamyurek_single-photon-level_2019} 
and separating the wave functions \cite{riedl_bose-einstein_2012} 
of the different hyperfine states during storage 
can be a major limiting factor of storage time (see also \cite{heshami_quantum_2016}).

After the signal pulse absorption, 
depending on the angle $\theta$ between signal and control pulse, 
the wavevector stored in the spin wave ranges 
from  $|\vec{k}_\text{c}|-|\vec{k}_\text{s}|=\frac{\omega_{gs}}{c}\approx 1/\, $mm 
to $|\vec{k}_\text{c}|+|\vec{k}_\text{s}|\approx 10/\, \rm{\mu}$m 
with a corresponding phase grating in the atomic state 
which can be scrambled by atomic motion
even with individual atoms retaining their phase. 
 Fig.~\ref{fig:t_motion_decoh} shows the resulting decoherence time scales 
when assuming thermal motion to be ballistic.
As the use of a buffer gas can restrict the ballistic motion of the atoms \cite{ledbetterSpinexchangerelaxationfreeMagnetometryCs2008}, it is possible to soften this limitation of the achievable deflection angle $\varphi$ in rubidium vapors.
Also, this wavevector corresponds to an additional momentum 
in the wave function of the $\ket{s}$ states 
leading to added velocities ranging from 
$\hbar |\vec{\kappa}|/m_\text{Rb}\approx 0.1\,{\rm  nm/(ms)}$ 
to $ 10\,{\rm \mu m/(ms)}$ in rubidium. 
To maximize storage time, it might be advisable to start with a
manipulation $\vec{\delta}_1=-\vec{\kappa}$ right after the
absorption process, thus removing the phase grating and stored
momentum mentioned above. 
In set-ups where almost parallel signal and control pulses 
must be chosen to avoid large stored momenta $\kappa$, 
our method of manipulation could be chosen to relax this constraint.
Directly before the emission
process, the wavevector can be reintroduced to the spin wave
together with the intended total manipulation $\vec{\delta}$,
thus minimising the influence of atomic motion:
$\vec{\delta}_2=-\vec{\delta}_1+\vec{\delta}=\vec{\kappa'}$.
For estimating the temperature regimes at which different wavevectors 
can be created or compensated by the proposed method, the color coding 
in Fig.~\ref{fig:t_motion_decoh} indicates how the time scale 
for manipulation compares to the decoherence from ballistic thermal motion. 
As manipulation time $T$, 
the values shown in Fig.~\ref{fig:manipulationovervarphi} are used, 
while accounting for the finite rise time of the coil 
by an additional fixed duration $2t_{\text{rise}}=10\,\mu$s. 
The time scale for decoherence (cf.~\cite{zhao_long-lived_2009})
$t_{\text{decoh}}$ is estimated by
the time it takes an atom at thermal velocity $v_{\text{th}}$ to 
traverse a significant fraction of the spin wave phase grating 
given by $\kappa$: $t_{\text{decoh}}=1/(v_{\text{th}}\kappa)$, 
where $v_{\text{th}}=\sqrt{k_B T_{\text{Rb}}/m_{\text{Rb}}}$, 
with $k_B$ being Boltzmann's constant,  
$T_{\text{Rb}}$ the temperature of the rubidium ensemble 
and $m_{\text{Rb}}$ the atomic mass of rubidium.

\begin{figure}
	\centering
	\includegraphics[width=0.6\linewidth]{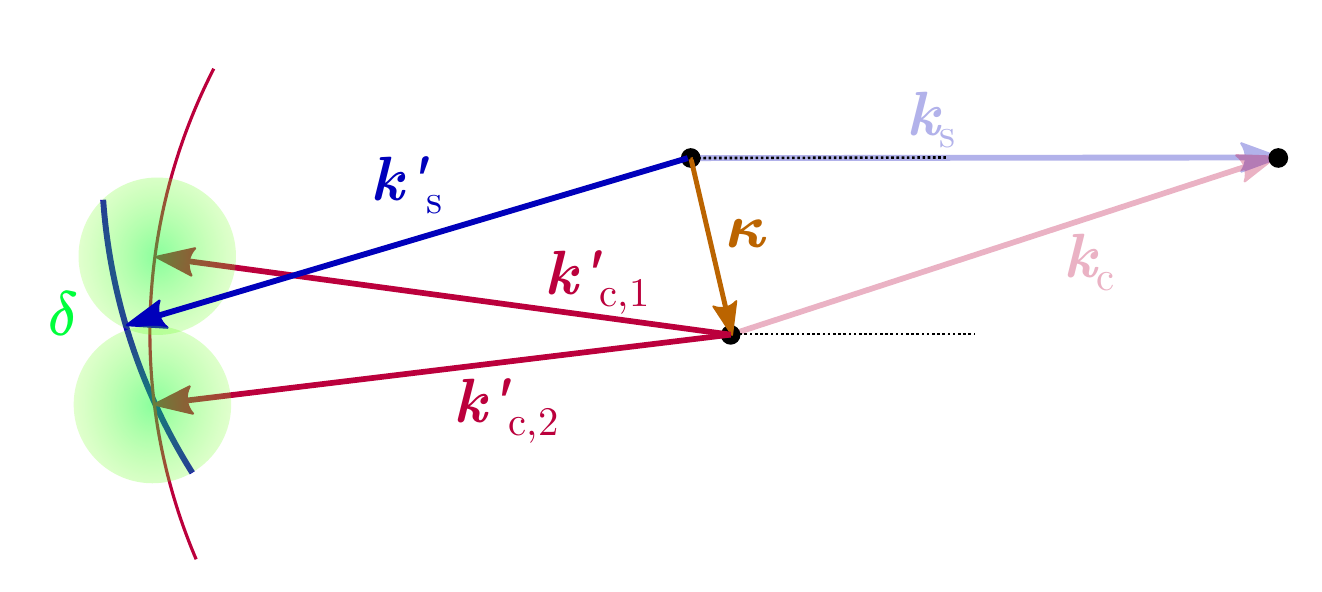}
	\caption{Drawing of the phase matching condition 
			when preparing multiple possible control pulses 
			and a moderate manipulation $|\vec{\delta}|\ll |\vec{k}_\text{s}|$. 
			The thick blue circle segment indicates the reachable emission directions 
			for the signal pulse.}
	\label{fig:beamer_config}
\end{figure}

When using dopants in solid bodies as active atomic ensemble, 
the decoherence due to ballistic motion is eliminated, 
such that even anti-parallel control and signal pulses 
($\theta= 180\,^\circ$) do not negatively affect 
the storage time (cf.~\cite{longdell_stopped_2005}). 
The solid medium will rescale the wave vectors involved, 
but the new time scales for shifting the spinwave wavevector 
will remain of the order of magnitude of a few to $10^2\,\mu$s, 
such that shifting the emission direction to 
arbitrary angles becomes possible. 
The condition for our Zeeman manipulation scheme to be applicable 
is a relative change in energy between the ground 
and storage states $g,e$ when applying an additional magnetic field. 
The estimated manipulation times assume a magnetic susceptibility 
corresponding to an electronic spin transition. 
Although the motional state of Bose-Einstein-condensates is outside the scope 
of our ansatz \eqref{eq:n}, existing experiments \cite{riedl_bose-einstein_2012, fleischhauer_electromagnetically_2005} 
indicate that photon storage can be described in a similar manner, 
and that due to the lack of thermal motion, 
the decoherence time of the spinwave is also less susceptible to its wavevector.
 This lets us expect that in BECs also, arbitrary deflection angles are achievable. 
In systems where only finite deflection angles can be achieved, 
the covered range of possible deflection angles 
can be increased by combining a fixed number of possible directions 
for the control pulse in emission with our proposed manipulation scheme 
as indicated in Fig.~\ref{fig:beamer_config}. 
        
\section{Summary}
\label{sec:Summary}

Using a fully three dimensional treatment, we regarded the possibilities of storing 
weak coherent or 
single-photon signal pulses 
in an atomic cloud of three-level-atoms and re-emitting
them in a controlled way in a new direction. 
The absorption of a photon in an ensemble of atoms results in a
spin-wave with well defined wavevector $\vec{\kappa}$ and envelope $S(\vec{r})$. 
The envelope influences emission efficiency and the shape of the re-emitted pulse,
whereas the wavevector reflects the 
momentum and energy-balance of two-photon absorption, 
with one photon from the signal
beam and one from the control beam.  We have shown that during 
storage the wavevector of the spin-wave can be modified by
e.g.~applying a magnetic field gradient,
without otherwise affecting the spin-wave. 
This modifies the momentum balance when 
the control beam is switched back on for
re-emitting the signal pulse in such a way that even without changing
the control beam a new emission direction can be selected.
In solid-state-based quantum memories arbitrary 
in-plane deflection angles can be achieved with 
reasonable coils and power-supplies. 
We expect that BECs, too, allow for arbitrary deflection angles.
In cold
atomic clouds or hot atomic vapors, due to atomic motion scrambling the phases of spin
waves that carry a significant wave vector, the resulting decoherence times are shortened, and correspondingly with the same coils and power-supplies
deflection angles are restricted to $\sim20\,$mrad
and $\sim0.2\,$mrad, respectively. This limitation can be softened
by restricting the thermal motion of the vapor with the use of a buffer gas.
This still allows for fast and efficient routing 
of photons into different beams or optical fibers.
Our numerical simulations show that the efficiency
of the whole process as
measured by the ratio of the emitted energy compared to the energy in
the incoming signal pulse is only moderately 
reduced for a beam emitted
in an arbitrary direction compared to the beam re-emitted in the 
direction of the incoming pulse, even without adjusting any other parameters. 
Here, the envelope of the spin wave 
with regards to the new emission direction is the limiting property for the efficiency.
Alternatively, one can also change the direction of the control beam
in order to send out the stored excitation in another direction, or
both methods can be combined.

The phases of the spin-wave are defined in
Hilbert space, i.e.~they control the coherent superposition of
many-particle states with
excitations localized at different positions in the atomic cloud whose
phase they define relative to the corresponding atomic ground
states. The effect that we described here is hence another remarkable
example of the phenomenon that phases in Hilbert space have impact on the interference
and propagation of photons in real configuration space, of which
quantum optics is full (see \cite{fabre_modes_2019} for a recent review). Using
the same control beam for emission as for absorption has the charm of
needing no movable elements 
such as micro-mirrors for deflecting the signal beam, and allows for
fast all-electronic control (on a time-scale 
on the order of few to $\sim100$ microseconds 
with reasonable magnetic field gradients, 
depending on the deflection angle)
of the emission direction, opening the path
for numerous applications of single-photon routing, such as
photon-multiplexing, quantum
communication to several parties, etc.
Due to the linearity of the dynamics, we expect that quantum superpositions 
of photons in different modes 
(e.g.~in different time bins, as commonly used in quantum memories)
will be propagated and re-directed with comparable efficiency 
as the pulses in a single mode considered here, 
but more work will be required to prove this.

The possibility of purely shifting the momentum stored in the spin-wave 
also in emission direction promises the possibility to assist 
existing ac-Stark effect based spin-wave manipulation methods by allowing
spin-wave multiplexing without any intrinsic loss 
introduced by non-linear phase factors. 

The scheme studied here focuses on deflection in the $xy$-plane.         
Slight deviations of the wavevector of the emitted light from 
the $xy$-plane should also be achievable, 
but deflection into arbitrary directions in the $4\pi$ spatial angle 
would need a rotation of the polarization vector as well. 
Alternatively, one might envisage a two-step deflection with the one 
in the $xy$-plane followed by another one in a plane perpendicular to it 
containing the wave-vector after the first deflection.  
In future works, it might be of interest to explore the proposed manipulation scheme in situations with further effects such as inhomogeneous broadening \cite{gorshkov_photon_2007-3}, exact atomic positions \cite{asenjo-garcia_exponential_2017,manzoni_optimization_2018}, and atomic interactions \cite{petrosyanCollectiveEmissionPhotons2021a}.

\clearpage

\bibliographystyle{unsrt}
\bibliography{quantum_router}

\clearpage
\appendix

\section{Coil properties} \label{seq:coil_properties}
\paragraph*{Goal parameters.} As example parameters for our atomic cloud 
we assume a spherical volume with $V=L^3=1\,$cm$^3$, 
implying a radius of $r\approx 0.6\,$cm. 
For the magnetic gradient coils, we assumed a magnetic gradient with $50\,$G/cm 
that can be ramped up or down in the order of $5\,\mu$s 
that extends over the whole of the atomic cloud.

\paragraph*{Corresponding coil parameters.} 
For a simple estimation of the necessary experimental current source 
and coil parameters, we assume the gradient coil to be a Maxwell coil pair with coil radius $a$. 
The rise time $\tau$ of the gradient coil is calculated as\cite{hidalgo-tobon_theory_2010}
\begin{align}
	\tau = \dfrac{L_{\text{c}}I}{V_{\text{c}}-RI} \overset{V_{\text{c}}\gg RI}{\approx} \dfrac{L_{\text{c}}I}{V_{\text{c}}}\overset{!}{=} 5\,\mu\text{s},\label{eq:time}
\end{align}
where $L_{\text{c}}$ is the inductance of the gradient coil, 
$R$ is its Ohmic resistance and $I$ is the equilibrium current 
flowing through the coil at applied voltage $V_{\text{c}}$.

The gradient created is given by
\begin{align}
	G = &\eta I \overset{!}{=} 50\,\text{G/cm}\label{eq:grad}\\
	&\eta \approx 0.64\ \mu_0 \dfrac{N_{\text{c}}}{a^2},\nonumber
\end{align}
where $\eta$ is the gradient coil efficiency, $\mu_0$ is the magnetic vacuum permittivity, 
and $N_{\text{c}}$ is the winding number of each coil.

The inductance of the Maxwell coils is approximated as
\begin{align}
	L_{\text{c}}\lessapprox 2 N_{\text{c}}^2 \pi a^2 \mu_0/(l+a/1.1) \approx \pi N_{\text{c}}^2 a \mu_0,\label{eq:inductance}
\end{align}
where the individual coil length $l$ was assumed to be $l=(1+0.1/1.1)a$.

Using a coil radius of $a=1\,$cm, we can solve \eqref{eq:grad} for $N_{\text{c}}I$, giving
\begin{align}
	50\,\text{G/cm}&\overset{!}{=}0.64 \mu_0 N_{\text{c}}I/a^2 \nonumber\\\Rightarrow N_{\text{c}}I &\overset{!}{=} a^2/(0.64\mu_0)\ G\approx 62.2\,\text{A}.
\end{align}
Inserting \eqref{eq:inductance} into \eqref{eq:time}, we get
\begin{align}
	5\,\mu\text{s}&\overset{!}{=} \dfrac{L_{\text{c}}I}{V_{\text{c}}} = \pi (N_{\text{c}}I) N_{\text{c}} a \mu_0 /V_{\text{c}} = \dfrac{\pi}{0.64} \dfrac{a^3 N_{\text{c}}G}{V_{\text{c}}}\nonumber \\\Rightarrow V_{\text{c}} &\overset{!}{=} \dfrac{G}{\tau} \dfrac{\pi a^3}{0.64} N_{\text{c}} \approx N_{\text{c}} \times 0.49\,\text{V}.
\end{align}
Choosing $N_{\text{c}}=63$, this gives
\begin{align*}
	V_{\text{c}}\approx 31\,\text{V} \quad \text{and} \quad I\approx 1\,\text{A}
\end{align*}
as solutions. The actual voltage needs to be increased by $RI$ 
to compensate for the coil's resistance.

\section{Time scale for adiabaticity.}
\label{sec:adiabaticity}
For the regarded Rb-87, the Hamiltonian of the ground state spin manifold 
under an external magnetic field $\vec{B}$ is given by
\begin{align}
	\hat{H}=A_{\text{HFS}}/\hbar^2 \hat{\vec{I}}\cdot\hat{\vec{J}} + \dfrac{\mu_{\text{Bohr}}}{\hbar} (g_S \hat{\vec{S}} + g_L \hat{\vec{L}})\cdot \vec{B},
\end{align}
where $A_{\text{HFS}}\approx h\ 3.42\,$GHz is the hyperfine coupling, 
$\mu_{\text{Bohr}}$ is the Bohr-magneton, $g_S\approx 2$ is the electron g-factor, 
$g_I\approx -0.001$ is the nuclear g-factor 
and $\vec{B}=(B_0+B_1(t))\vec{e}_z$ is the applied magnetic field. 
In our case, we have $L=0$, $J=S=1/2$ and $I=3/2$. 
In the $\ket{m_I}_z\otimes \ket{m_S}_z$-basis, 
the magnetic coupling is diagonal and the hyperfine coupling takes the form
\begin{align} 
	A_{\text{HFS}} \hat{\vec{I}}\cdot\hat{\vec{S}}/\hbar^2 = A_{\text{HFS}} \pmqty{3/4 & 0 & 0 & 0 & 0 & 0 & 0 & 0 \\ 
		0	& -3/4 & \sqrt{3}/2 & 0 & 0 & 0 & 0 & 0\\ 0 & \sqrt{3}/2 & 1/4 & 0 & 0 & 0 & 0 & 0 \\ 0 & 0 & 0 & -1/4 & 1 & 0 & 0 & 0\\ 0 & 0 & 0 & 1 & -1/4 & 0 & 0 & 0 \\ 0 & 0 & 0 & 0 & 0 & 1/4 & \sqrt{3}/2 & 0 \\
		0 & 0 & 0 & 0 & 0 & \sqrt{3}/2 & -3/4 & 0 \\ 0 & 0 & 0 & 0 & 0 & 0 & 0 & 3/4},
\end{align}
such that any state decay due to fluctuating magnetic fields 
only affects the 2-dimensional state subspaces described by
\begin{align}
	\ket{m_I=+3/2}\otimes \ket{m_S=-1/2} &\leftrightarrow \ket{m_I=+1/2}\otimes \ket{m_S=+1/2},\label{eq:H1} \\
	\ket{m_I=+1/2}\otimes \ket{m_S=-1/2} &\leftrightarrow \ket{m_I=-1/2}\otimes \ket{m_S=+1/2},\label{eq:H2} \\
	\ket{m_I=-1/2}\otimes \ket{m_S=-1/2} &\leftrightarrow \ket{m_I=-3/2}\otimes \ket{m_S=+1/2}. \label{eq:H3}
\end{align}

Thus the question of adiabaticity for ramping up the B1-field 
can be regarded separately for the two-level subsystems. 

The corresponding two-level Hamiltonians (up to a two-level-global 
energy shift for zero-averaged eigenvalues) become
\begin{align}
	\eqref{eq:H1} &\leftrightarrow \hat{H} = A_{\text{HFS}} \pmqty{-1/2 & \sqrt{3}/2 \\ \sqrt{3}/2 & 1/2} + \mu_{\text{Bohr}} \pmqty{\frac{g_S-g_I}{2} & 0 \\ 0 & \frac{-g_S+g_I}{2}} (B_0+B_1(t)) \\
	\eqref{eq:H2} &\leftrightarrow \hat{H} = A_{\text{HFS}} \pmqty{0 & 1 \\ 1 & 0} + \mu_{\text{Bohr}} \pmqty{\frac{-g_S+g_I}{2} & 0 \\ 0 & \frac{g_S-g_I}{2}} (B_0+B_1(t)) \\
	\eqref{eq:H3} &\leftrightarrow \hat{H} = A_{\text{HFS}} \pmqty{1/2 & \sqrt{3}/2 \\ \sqrt{3}/2 & -1/2} + \mu_{\text{Bohr}} \pmqty{\frac{-g_S+g_I}{2} & 0 \\ 0 & \frac{g_S-g_I}{2}} (B_0+B_1(t)),
\end{align}

which, assuming $B_1(t)=B_1\ t/\tau$ to be linear in time,  all have the form 
\begin{align}
	\hat{H} = \hbar v \hat{\sigma}_x + \hbar(\epsilon +  b t) \hat{\sigma}_z,\label{H_general}
\end{align}
with the corresponding values
\begin{align}
	\hbar v=\left\{ \mqty{\sqrt{3}/2 A_{\text{HFS}}\\  A_{\text{HFS}}\\ \sqrt{3}/2 A_{\text{HFS}}} \right. , \ 
	\hbar \epsilon =\left\{ \mqty{-A_{\text{HFS}}/2+\frac{g_S-g_I}{2}B_0 \\
		\frac{-g_S+g_I}{2}B_0 \\
		A_{\text{HFS}}/2+\frac{-g_S+g_I}{2}B_0 }\right. ,\  
	\hbar b =\left\{ \mqty{\mu_{\text{Bohr}}\frac{g_S-g_I}{2}B_1/\tau & \text{for }\eqref{eq:H1},\\
		\mu_{\text{Bohr}}\frac{-g_S+g_I}{2}B_1/\tau & \text{for }\eqref{eq:H2},\\
		\mu_{\text{Bohr}}\frac{-g_S+g_I}{2}B_1/\tau & \text{for }\eqref{eq:H3}.}\right.
\end{align}

The actual situation is described by $t\in[0,\tau]$, 
but extending this to $\pm\infty$ allows 
for an analytical solution of the transition probability $p$ 
using the Landau-Zener formula\cite{brundobler_s-matrix_1993}
\begin{align}
	p = e^{-\pi z}, \qquad z = \dfrac{|v|^2}{|2b|},
\end{align}
where $z$ is the Landau-Zener parameter. With $B_1\le 500\,$G 
throughout our atomic ensemble and $\tau\gtrsim 1\,\mu$s, we have
\begin{align}\label{eq:landau-zener-parameter}
	z \ge (A_{\text{HFS}})^2/|2\hbar \mu_{\text{Bohr}}\dfrac{g_S-g_I}{2}B_1/\tau | \approx 50 \dfrac{\tau}{1\,\text{ns}} \overset{\tau\gtrsim 1\,\mu\text{s}}{\gg} 1,
\end{align}
such that state-transitions due to ramping up the B1-field can be neglected.

\begin{figure}
	\centering
	\includegraphics[width=.7\linewidth]{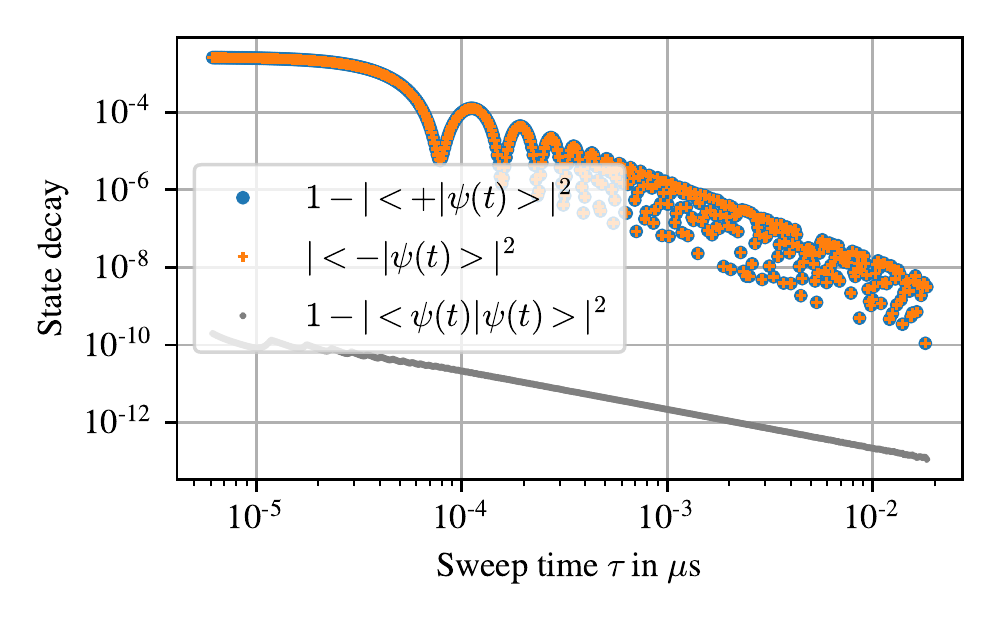}
	\caption{\label{fig:num_adiabaticity} 
		Numerical results for the final state after $B_1$ has been swept 
		from zero to its full value ($B_1(\tau)=250\,$G) in time $\tau$. 
		The initial state is taken as $\ket{+}$ 
		and the parameters are taken from \eqref{eq:H1}.}
\end{figure}

\begin{figure}
	\centering
	\includegraphics[width=.7\linewidth]{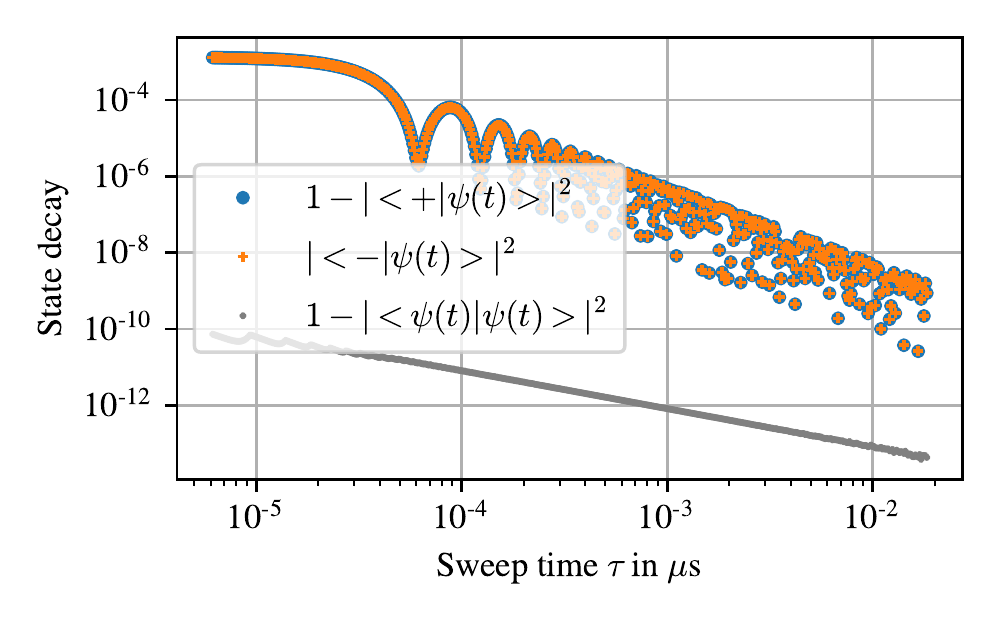}
	\includegraphics[width=.7\linewidth]{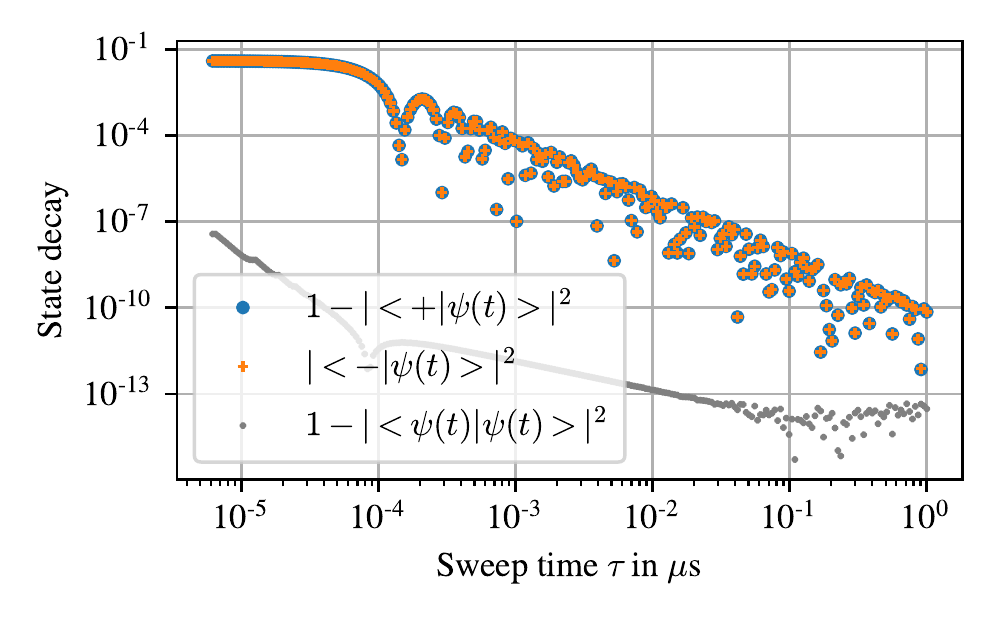}
	\caption{\label{fig:num_adiabaticity2} 
	Numerical results for the final state after $B_1$ has been swept 
	from zero to its full value ($B_1(\tau)=250\,$G) in time $\tau$. 
	The initial state is taken as $\ket{+}$ and the parameters 
	are taken from \eqref{eq:H2}, 
	while the lower plot additionally sets $B_0$ to zero 
	to check whether this offset field is necessary for adiabaticity.}
\end{figure}

\paragraph*{Finite times.} As the assumption of $t\in (-\infty,\infty)$ 
is not fulfilled in experiment, we do a numerical integration 
of the time evolution in order to make sure 
that $\tau\gtrsim1\,\mu$s is a safe regime 
with regards to negligible disturbance of the state.
We choose the two-level Hamiltonians as introduced above 
in \eqref{H_general}. As \eqref{eq:H1} and \eqref{eq:H3} are equivalent 
except for exchanging the two basis states, 
only the parameters for \eqref{eq:H1} and \eqref{eq:H2} are regarded separately. 
The numerical integration evolves the state from $t=0$ to $t=\tau$, 
using the instantaneous eigenbasis, and $\ket{+}:=(1,0)^T$ as initial state. 
A third calculation is made with parameters from \eqref{eq:H2}, but with $B_0=0$, 
to test whether the background field that was introduced 
for approximately linearising the response to $B_1$ is also necessary to achieve adiabaticity. 
Figures \ref{fig:num_adiabaticity} and \ref{fig:num_adiabaticity2} show the results, 
which clearly indicate that adiabaticity remains a good approximation 
for $\tau$ in the $\mu$s regime even in the case of $t\in[0, \tau]$. Comparing these results with those of Appendix \ref{seq:coil_properties} we find that the speed with which the magnetic field can be altered is in practice limited by technical limitations while the fundamental limitations from the adiabaticity condition become relevant only at time scales that are more comparable to the hyperfine interaction as indicated by the Landau-Zener paremeter calculated in \eqref{eq:landau-zener-parameter}.


\end{document}